\documentstyle[psfig]{mn}

\newif\ifAMStwofonts

\ifoldfss
  \ifCUPmtlplainloaded \else
    \NewTextAlphabet{textbfit} {cmbxti10} {}
    \NewTextAlphabet{textbfss} {cmssbx10} {}
    \NewMathAlphabet{mathbfit} {cmbxti10} {} 
    \NewMathAlphabet{mathbfss} {cmssbx10} {} 
  \fi
  \ifAMStwofonts
    \ifCUPmtlplainloaded \else
      \NewSymbolFont{upmath} {eurm10}
      \NewSymbolFont{AMSa} {msam10}
      \NewMathSymbol{\upi}     {0}{upmath}{19}
      \NewMathSymbol{\umu}     {0}{upmath}{16}
      \NewMathSymbol{\upartial}{0}{upmath}{40}
      \NewMathSymbol{\leqslant}{3}{AMSa}{36}
      \NewMathSymbol{\geqslant}{3}{AMSa}{3E}

       \let\le=\leqslant
       \let\ge=\geqslant
    \fi
  \fi
\fi 

\ifnfssone
  \newmathalphabet{\mathit}
  \addtoversion{normal}{\mathit}{cmr}{m}{it}
  \addtoversion{bold}{\mathit}{cmr}{bx}{it}
  \newmathalphabet{\mathbfit} 
  \addtoversion{normal}{\mathbfit}{cmr}{bx}{it}
  \addtoversion{bold}{\mathbfit}{cmr}{bx}{it}
  \newmathalphabet{\mathbfss} 
  \addtoversion{normal}{\mathbfss}{cmss}{bx}{n}
  \addtoversion{bold}{\mathbfss}{cmss}{bx}{n}
  \ifAMStwofonts
    \ifCUPmtlplainloaded \else
      \UseAMStwoboldmath
      \makeatletter
      \new@mathgroup\upmath@group
      \define@mathgroup\mv@normal\upmath@group{eur}{m}{n}
      \define@mathgroup\mv@bold\upmath@group{eur}{b}{n}
      \edef\UPM{\hexnumber\upmath@group}
      \new@mathgroup\amsa@group
      \define@mathgroup\mv@normal\amsa@group{msa}{m}{n}
      \define@mathgroup\mv@bold\amsa@group{msa}{m}{n}
      \edef\AMSa{\hexnumber\amsa@group}
      \makeatother
      \mathchardef\upi="0\UPM19
      \mathchardef\umu="0\UPM16
      \mathchardef\upartial="0\UPM40
      \mathchardef\leqslant="3\AMSa36
      \mathchardef\geqslant="3\AMSa3E

       \let\le=\leqslant
       \let\ge=\geqslant
    \fi
  \fi
\fi 

\ifnfsstwo
  \DeclareMathAlphabet{\mathbfit}{OT1}{cmr}{bx}{it}
  \SetMathAlphabet\mathbfit{bold}{OT1}{cmr}{bx}{it}
  \DeclareMathAlphabet{\mathbfss}{OT1}{cmss}{bx}{n}
  \SetMathAlphabet\mathbfss{bold}{OT1}{cmss}{bx}{n}
  \ifAMStwofonts
    \ifCUPmtlplainloaded \else
      \DeclareSymbolFont{UPM}{U}{eur}{m}{n}
      \SetSymbolFont{UPM}{bold}{U}{eur}{b}{n}
      \DeclareSymbolFont{AMSa}{U}{msa}{m}{n}
      \DeclareMathSymbol{\upi}{0}{UPM}{"19}
      \DeclareMathSymbol{\umu}{0}{UPM}{"16}
      \DeclareMathSymbol{\upartial}{0}{UPM}{"40}
      \DeclareMathSymbol{\leqslant}{3}{AMSa}{"36}
      \DeclareMathSymbol{\geqslant}{3}{AMSa}{"3E}

       \let\le=\leqslant
       \let\ge=\geqslant
    \fi
  \fi
\fi 

\ifCUPmtlplainloaded \else
  \ifAMStwofonts \else 
    \def\upi{\pi}
    \def\umu{\mu}
    \def\upartial{\partial}
  \fi
\fi

\title{$UBVRI$ CCD photometric study of the open clusters Basel 4 and NGC 7067}
\author[R. K. S. Yadav and Ram Sagar]
       {R. K. S. Yadav$^{1}$\thanks{E-mail: rkant@iucaa.ernet.in} and Ram Sagar$^{2}$\thanks{E-mail: sagar@upso.ernet.in}\\
        $^{1}$Inter-University Centre for Astronomy and Astrophysics, Ganeshkhind, 
	Pune 411 007, India\\
        $^{2}$State Observatory, Manora Peak Nainital 263 129, India}
\date{Accepted ---------.
      Received ---------;
      }

\pagerange{\pageref{firstpage}--\pageref{lastpage}}
\pubyear{2003}
\begin{document}
\maketitle
\label{firstpage}
\begin{abstract}
In this paper we present $UBVRI$ CCD photometry in the region of two young open star 
clusters Basel 4 and NGC 7067 for the first time. Our sample consists of $\sim$ 4000 
stars down to $V$ $\sim$ 21 mag. Stellar surface density profile indicates 
that radius of Basel 4 and NGC 7067 are about 1.8 and 3.0 arcmin respectively. The $(U-B)$
versus $(B-V)$ diagrams indicate that metallicity of NGC 7067 is solar while that of
Basel 4 is $Z \sim$ 0.008. We estimate the mean value of $E(B-V)$ = 0.45$\pm$0.05 and 
0.75$\pm$0.05 mag for Basel 4 and NGC 7067 
respectively. The analysis of 2MASS $JHK$ data in combination with the optical data 
in both the clusters yields $E(J-K)$ = 0.30$\pm$0.20 mag and $E(V-K)$ = 1.60$\pm$0.20 
mag for Basel 4 while $E(J-K)$ = 0.40$\pm$0.20 mag and $E(V-K)$ = 2.10$\pm$0.20 mag 
for NGC 7067. Furthermore, colour excess diagrams show a normal interstellar 
extinction law towards both the clusters.

Using the intrinsic colour-magnitude diagrams of the cluster members, we estimated the 
distances of the clusters as 3.0$\pm$0.2 and 3.6$\pm$0.2 Kpc for Basel 4 and NGC 7067
respectively. By fitting the proper metallicity isochrones to 
the bright cluster members we estimated the age of the clusters as 200$\pm$50 and 100$\pm$25 
Myr for Basel 4 and NGC 7067 respectively. The mass function slope which is derived by 
applying the corrections of field star contamination and data incompleteness 
are $1.55\pm0.25$ and $1.68\pm0.47$ for Basel 4 and NGC 7067 respectively. The 
values of mass function slopes are thus not too different from the Salpeter's (1955) value. 
 Mass segregation is observed in both the clusters which may be due to the
dynamical evolutions or imprint of star formation processes or both.\\

\end{abstract}


\section{Introduction}

Open clusters are ideal objects for the study of Galactic disk. The young 
open clusters are used to determine spiral
 arm structure, to investigate the mechanisms of star formation and its 
recent history, and to constrain the initial luminosity and mass function 
in aggregates of stars etc. For such studies, it is important to know the basic 
parameters of the clusters. The Colour-Magnitude (CM) and Colour-Colour (CC) 
diagrams of an open cluster are valuable tools for obtaining their basic informations,  
such as its distance and age, and for studying both 
interstellar extinction in the direction of cluster and stellar evolution.
 An important aspect for understanding star formation and stellar evolution 
is the question: how many stars of which masses formed or exist in an 
ensemble of stars ? A function which described the frequency distribution of 
stellar masses is called the stellar mass function. Star clusters are suitable 
objects for mass function determination as members formed (more or less) 
at the same time and from the same cloud. In addition to this, study of mass 
segregation in open clusters provides a clue about the spatial distribution of high 
and low mass stars within the clusters. Generally, it is found that high mass 
stars are concentrated towards the center of the clusters in comparison to 
lower mass stars. The 
cause of such kind of distribution is still not well understood whether it 
is due to dynamical evolution or imprint of star formation itself.

\begin{table*}
 \centering
 \begin{minipage}{140mm}
 \caption{General information about the clusters under study taken from 
Dias et al. (2002)}
\begin{tabular}{|c|ccccccccc|}
\hline
Cluster&IAU&OCL&l&b&Trumpler&Radius&Distance&$E(B-V)$&log(age)\\
&&&(deg)&(deg)&class&(arcmin)&(Kpc)&(mag)&(yrs)\\
\hline
Basel 4&C0545+302&455&179.23&1.20&II 1p&2.5&5.6&0.53&7.0\\
NGC 7067&C2122+478&208&91.19&-1.67&II 1p&1.5&1.3&0.85&7.5\\
\hline
\end{tabular}
\end{minipage}
\end{table*}

In the light of above discussions, we conducted $UBVRI$ CCD stellar photometry 
in two young open star clusters Basel 4 and NGC 7067 aiming to investigate the 
cluster's basic parameters (e.g. reddening, distance and age), mass function 
and mass segregation etc. The existing basic informations on both the clusters 
are given in Table 1. 
The plan of the paper is as follows. In Sec. 2 we summarize 
the previous studies of Basel 4 and NGC 7067, while Sec. 3 is dedicated on 
the observation and data reduction strategies. Sec. 4 deals with the determination 
of clusters basic parameters as well as detail study of interstellar extinction, 
mass function and mass segregation in the clusters under study. Finally, Sec. 5 
summarizes our findings.

\section[]{Earlier investigations}
{\bf Basel 4}: This cluster was studied by Svolopoulos
(1965) photographically first in RGU system. According to him the location of this cluster 
coincides with spiral arm +III which could be expected $-$ if existing at all
 $-$ at a similar distance. In any case, it is remarkable that typical
representatives of the galactic disk population are located so far out in the
direction of the galactic anticenter. He classified this cluster as a III 2m. In addition
 to this, he also concluded that Basel 4 is 2$\times$10$^{7}$ yrs old, and has total
apparent diameter of 6$^{\prime}$.4 at a distance of 5.9. To our
knowledge no other studies have been carried out so far.\\ 

\noindent {\bf NGC 7067}: This cluster was first studied by Becker (1963). It is a poor
young open cluster lying in Cygnus spiral arm. It was again revisited by
Becker (1965) and indicated that the earliest spectral type of the cluster member
 is b0.5. He also estimated the cluster angular diameter of 2$^\prime$.1, which
corresponds to a linear diameter of 2.6 pc. Hassan (1973) also studied this
cluster photoelectrically and derived a distance of about 4.4 Kpc having $E(B-V) =
0.83$ mag and age less than $10^{7}$ years. Dias et al. (2002) mentioned a distance of 
1.3 Kpc for this cluster. The distance determination to the cluster is thus quite uncertain.

\section{Optical observations and data reductions}

\begin{figure*}
\begin{center}
\hbox{
\psfig{file=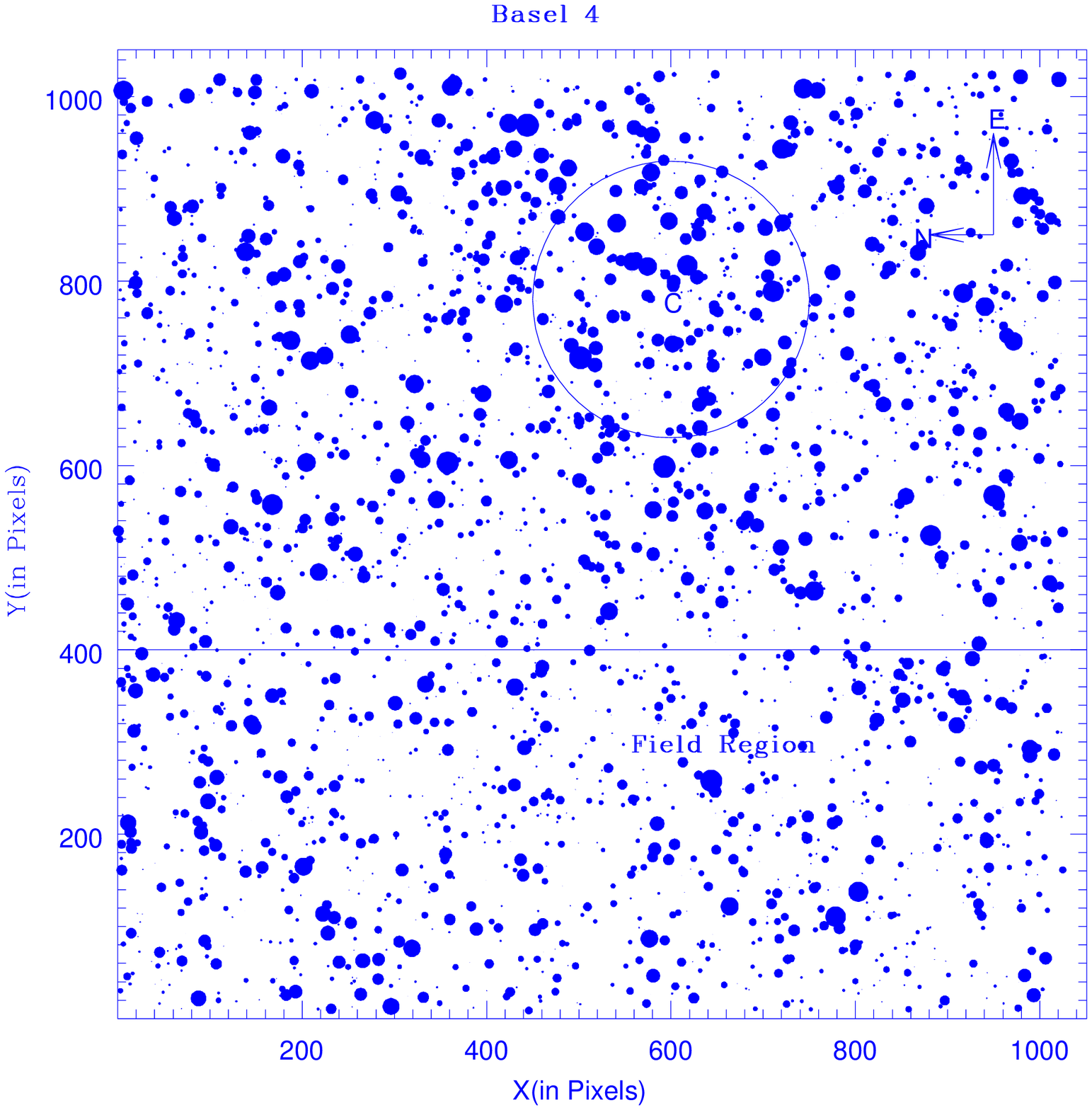,width=9cm,height=9cm}
\psfig{file=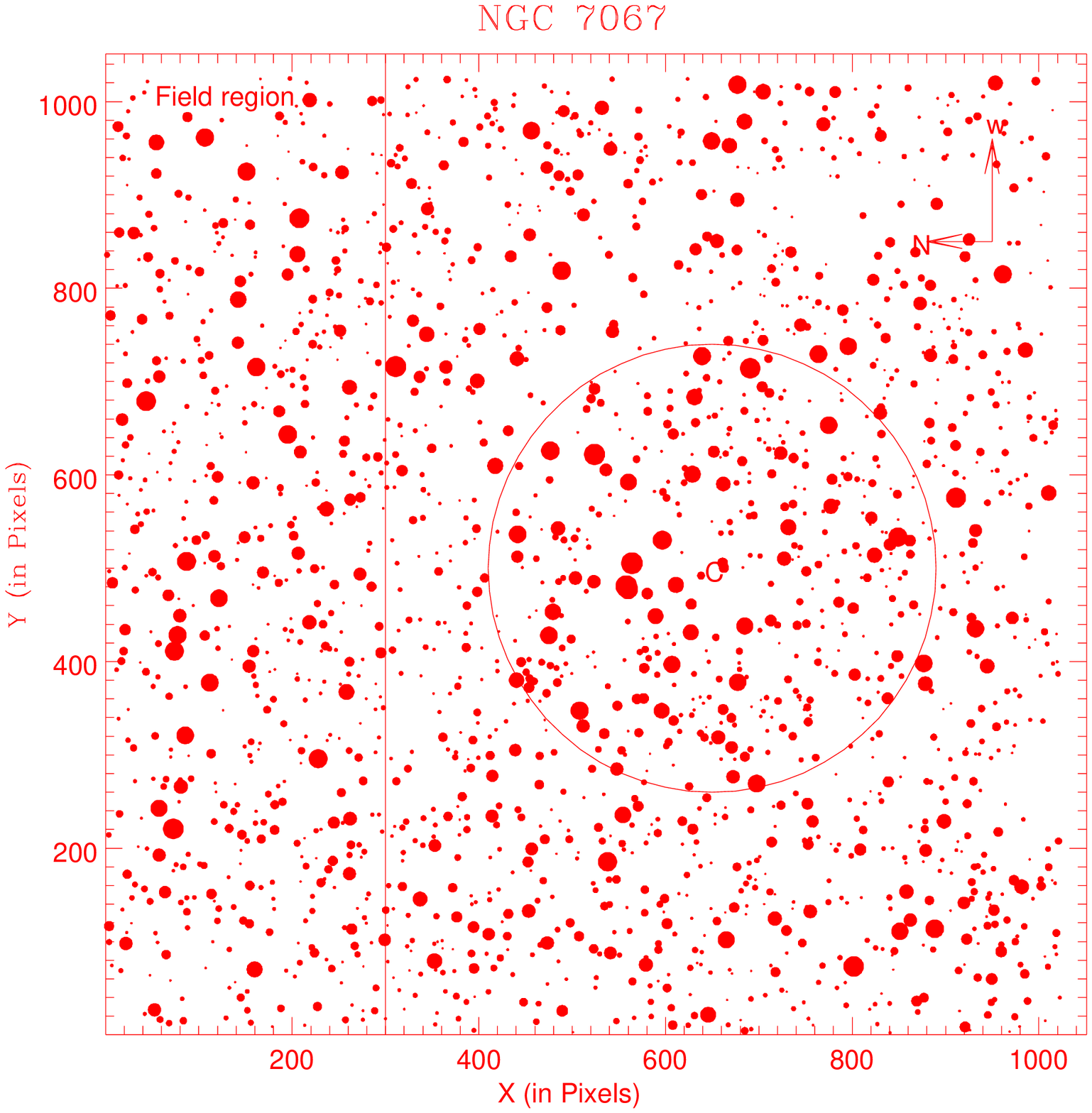,width=9cm,height=9cm}
}
\caption{Identification maps for the cluster and field regions of Basel 4
and NGC 7067. The (X, Y) coordinates are in pixel units corresponding
to 0$^{\prime\prime}$.72 on the sky. Direction is indicated in the corresponding 
map. Filled circles of different sizes represent brightness of the stars. Smallest size
denotes stars of $V$$\sim$21 mag. Open circles having centre at 'C' in 
the map represent the respective cluster size.}
\end{center}
\end{figure*}

\subsection {CCD Photometric observations}

We used CCD imaging to obtain $UBV$ Johnson and $RI$ Cousins photometry of the 
stars in the region of the open clusters Basel 4 and NGC 7067 on 02/03 Jan 2000 
and 11/12 Oct 2001 respectively. The data were obtained using 2K$\times$2K CCD 
system at the f/13 Cassegrain focus of the 104-cm Sampurnanand telescope 
of the State Observatory, Naini Tal. Log of CCD observations is given in Table 2.
 The 0$^{\prime\prime}$.36/pixel plate scale resulted in a field of view of 
12$^{\prime}$.3$\times$12$^{\prime}.3$. The read-out noise and gain of the CCD 
are 5.3 e$^{-}$ 
and 10 e$^{-}$/ADU respectively. For the accurate photometric measurements of 
fainter stars, 2 to 3 deep exposures were taken in each passband. Furthermore, 
observations were taken in 2$\times$2 pixel binning mode to improve the 
S/N ratio. An identification map of cluster and field regions for both the clusters 
are shown in Fig 1. Besides them, a number of standard star field 
were also observed for calibration purposes. We observed M67 (open cluster) and 
field PG0231$+$051 of Landolt (1992) for calibrating Basel 4 and NGC 7067 respectively.
 The $V$ mag range of stars used for calibration is 11$-$13 mag in M67 and 12$-$16 
mag in PG0231$+$051 while the $(V-I)$ colour range is 0.5$-$1.1 mag in M67 and 
$-$0.5$-$2.0 in PG0231$+$051. Thus, the standard stars in these fields provide a 
good magnitude and colour coverage, essential to obtain reliable photometric 
transformations. The standard field are also observed 
in $UBVRI$ at different airmasses to obtain a reliable estimate of the 
atmospheric extinction coefficients. For correcting the bias level to 
the image, a number of bias frame were taken during the observations while 
for the flat field correction, a number of flat frames were taken on the 
twilight sky in each filter. 

\begin{table*}
\centering
\begin{minipage}{150mm}
\caption{Log of CCD observations alongwith equatorial coordinates for the epoch 2000.}
\begin{center}
\begin{tabular}{cccc}
\hline
Region & Filter  &Exposure Time &Date\\
&&(in seconds)   & \\
\hline
Basel 4&$U$&1800$\times$2, 300$\times$1&02/03 Jan 2000\\
$\alpha_{2000}=07^{h}32^{m}00^{s}$&$B$&1200$\times$2, 180$\times$1&,,\\
$\delta_{2000}=+30^{d}12^{\prime}57^{\prime\prime}$&$V$&900$\times$2, 120$\times$1&,,\\
&$R$&240$\times$3, 60$\times$1&,,\\
&$I$&240$\times$3, 60$\times$1&,,\\
\hline
NGC 7067&$U$&1800$\times$3, 300$\times$2&11/12 Oct 2001\\
$\alpha_{2000}=21^{h}24^{m}11^{s}$&$B$&1200$\times$3, 240$\times$2&,,\\
$\delta_{2000}=+48^{d}00^{\prime}57^{\prime\prime}$&$V$&900$\times$3, 180$\times$2&,,\\
&$R$&600$\times$3, 120$\times$2&,,\\
&$I$&300$\times$3, 60$\times$2&,,\\
\hline
\end{tabular}
\end{center}
\end{minipage}
\end{table*}

\subsection{Data Reductions}

The CCD images were processed using IRAF data reduction package. Then, for a 
given filter, frame of the same exposure time were combined into one, to 
improve the statistics of the faintest stars. Instrumental magnitudes were 
derived through Point Spread Function (PSF) fitting using DAOPHOT (Stetson 1987) 
within MIDAS. During the process of determining PSF, we used several 
well isolated stars to construct a single PSF for the entire frame on each 
exposure. The bright stars were measured on the frames with short exposure 
times, as they were saturated in the longer exposure frames. 

For transforming the instrumental magnitude to the standard magnitude, the photometric 
calibration equations are as follows:\\

$u=U+6.40\pm0.01-(0.03\pm0.02)(U-B)+0.59X$

$b=B+4.39\pm0.01-(0.02\pm0.01)(B-V)+0.36X$

$v=V+4.08\pm0.01-(0.01\pm0.01)(B-V)+0.23X$

$r=R+4.00\pm0.01-(0.02\pm0.01)(V-R)+0.18X$

$i=I+4.55\pm0.01-(0.01\pm0.02)(R-I)+0.13X$\\

Where $U, B, V, R$ and $I$ are the standard magnitudes and $u, b, v, r$ and $i$ are 
the instrumental aperture magnitudes  
normalised for 1 second of exposure time and $X$ is the airmass. We have ignored the 
second order colour correction terms as they are generally small in comparison to 
other errors present in the photometric data reduction. The errors in zero point 
and colour coefficients are estimated from the deviation of data points from the 
linear relation and are $\sim$ 0.01 mag. Using these transformation equation, 
we calibrated the CCD instrumental magnitudes. For generating the local standards, 
we selected many well isolated stars in the observed region and used the 
{\bf DAOGROW} programme for the construction of an aperture growth curve required 
for determining the difference between aperture and profile fitting magnitudes. 
Table 3 gives the photometric errors as a function of 
magnitude. The internal errors estimated on the S/N ratio of the stars as output of 
the {\bf ALLSTAR} mainly produce the scatter in the various CC and CM diagrams of the clusters. 
It can be noticed that the errors become large ($\ge$ 0.1 mag) for stars fainter than 
$V$= 20 mag, so the measurements should be considered unreliable below this magnitude. The 
final photometric data is available in electronic form at the WEBDA site {\it http://obswww.unige.ch/webda/} and also from the authors. The format of the table is listed in Table 4 for Basel 4 and NGC 7067. 

\subsection{Comparison with previous photometric study}
As mentioned in Sec. 2, only NGC 7067 has photoelectric data given by Becker (1965) and we 
compared our data with this data. Table 5 presents average differences (in the 
sense: our values minus those of other) along with their standard deviations. The 
difference $\Delta$ in $V$, $(B-V)$ and $(U-B)$ are plotted in Fig 
2. Table 6 and Fig. 2 indicate that our $V$ magnitudes are systematically brighter 
by $\sim$ 0.05 mag without any dependence on the stellar magnitude. The $\Delta (B-V)$ 
values show a good agreement with the photoelectric data while $\Delta (U-B)$ show 
a decreasing trend with the $V$ mag. 

\begin{table}
 \centering
\caption{Internal photometric errors in magnitude as a function of brightness. $\sigma$ is 
the standard deviation per observation in magnitude.}

\begin{tabular}{cccccc}
\hline
Magnitude range& $\sigma$$_{U}$&$\sigma$$_{B}$&$\sigma$$_{V}$  &$\sigma$$_{R}$ &
$\sigma$$_{I}$\\
\hline
$\le$12.0&0.01&0.01&0.01&0.01&0.01\\
12.0 - 13.0&0.01&0.01&0.01&0.01&0.01\\
13.0 - 14.0&0.01&0.01&0.01&0.01&0.01\\
14.0 - 15.0&0.01&0.01&0.01&0.01&0.01\\
15.0 - 16.0&0.01&0.01&0.01&0.01&0.01\\
16.0 - 17.0&0.02&0.01&0.01&0.01&0.02\\
17.0 - 18.0&0.03&0.02&0.02&0.02&0.03\\
18.0 - 19.0&0.04&0.05&0.03&0.05&0.06\\
19.0 - 20.0&0.05&0.09&0.05&0.09&0.08\\
\hline
\end{tabular}
\end{table}

\subsection {Near - IR data}
We have used the Two Micron All Sky Survey (2MASS) $J$(1.25 $\mu$m), $H$(1.65 $\mu$m) and 
$K$$_{s}$(2.17 $\mu$m) data for both the clusters Basel 4 and NGC 7067. 2MASS data is 
taken from observations with two highly - automated 1.3-m telescopes one at Mt Hopkins, AZ, and 
second at Cerro Tolalo Inter-American Observatory (CTIO), Chile. The data are complete up to 
16.0 mag in $J$, 15.5 mag in $H$ and 15.0 in $K_s$. The uncertainty is 0.155 mag for a star 
of $K_s$ $\sim$ 16.5 mag. The $K_s$ magnitudes are converted into $K$ 
magnitude following Persson et al. (1998). 2MASS data is available at web site 
{\it http://www.ipac.caltech.edu/2MASS/}. 

\begin{figure}
\hspace{-8.0cm}\psfig{file=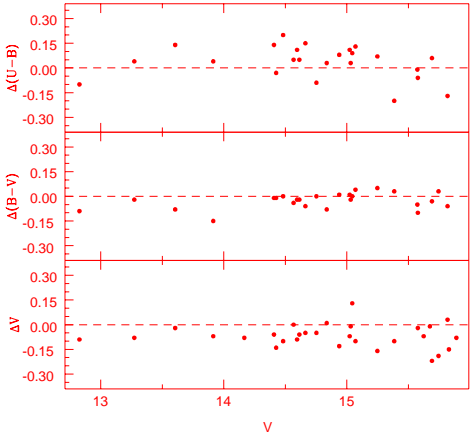, width=9cm,height=8cm}
\caption{Comparison of our photometry with photoelectric data of Becker (1965) 
for NGC 7067.}
\end{figure}

\begin{table*}
 \centering
 \begin{minipage}{140mm}
\caption{CCD relative (X, Y) positions and $V$, $(U-B)$, $(B-V)$, $(V-R)$ and $(V-I)$ 
photometric magnitudes of few stars, as a sample measured in the region of the 
cluster Basel 4 and NGC 7067. Stars are numbered in increasing order of X value. 
The last column represent the photometric membership informations where m and nm 
represent the member and non-member stars respectively.}

\begin{center}
\begin{tabular}{ccccccccc}
\hline
Star&X&Y&$V$&$(U-B)$&$(B-V)$&$(V-R)$&$(V-I)$&Mem\\
&(pixel)&(pixel)&(mag)&(mag)&(mag)&(mag)&(mag)\\
\hline
&&&&Basel 4&&&\\
1&     1.87&  701.86& 19.64&    * &   * &     0.69&  1.61&  nm\\
2&     2.01&  519.82& 18.76&    * &   * &     0.40&  1.17&  nm\\
3&     2.71&  790.32& 20.66&    * &   * &     0.48&  1.52&  nm\\
4&     2.87&   30.33& 19.63&    * &   * &     0.58&  1.21&  nm\\
5&     3.34&  877.65& 20.33&    * &   * &     0.33&  1.34&  nm\\
&&&&&&&\\
&&&&NGC 7067&&&\\
1&     1.69&    835.80&  19.37&      *&     *&   0.68&   1.61&  nm\\
2&     3.47&    214.41&  19.57&      *&     *&   0.83&   1.93&  nm\\
3&     3.87&    116.54&  17.60&      *&     *&   1.24&   2.91&  nm\\
4&     3.94&     99.46&  19.33&      *&     *&   0.88&   1.87&  nm\\
5&     4.01&    496.79&  20.22&      *&     *&   0.78&   1.70&  nm\\
\hline
\end{tabular}
\end{center}
\end{minipage}
\end{table*}

\begin{table*}
 \begin{minipage}{140mm}
 \caption{Comparison of our photometry with Becker (1965) for the cluster 
NGC 7067. The difference ($\Delta$) is always in the sense present minus 
comparison data. The mean and standard deviations in magnitude are based on 
N stars. Few deviated points are not included in the average determination.}
\begin{tabular}{cccccc}
\hline
Cluster&Comparison data&$V$ range&$<\Delta$$V>$&$<\Delta(B-V)$$>$&$<\Delta(U-B)>$\\
&&&Mean$\pm$$\sigma$(N)&Mean$\pm$$\sigma$(N)&Mean$\pm$$\sigma$(N)\\
\hline
NGC 7067&Becker (1965)&$<$ 14.0&\hspace{-0.15cm}0.06$\pm$0.03(4)&$-$0.10$\pm$0.03(4)&0.07$\pm$0.14(4)\\
&&14.0 $-$ 15.0&0.06$\pm$0.04(11)&$-$0.02$\pm$0.02(10)&$-$0.07$\pm$0.08(10)\\
&&15.0 $-$ 16.0&$-$0.07$\pm$0.09(14)&$-$0.006$\pm$0.04(10)&0.06$\pm$0.12(9)\\
\hline
\end{tabular}
\end{minipage}
\end{table*}

\section {Analysis of the data}

\subsection {Cluster radius and radial stellar surface density}

\begin{figure}
\psfig{figure=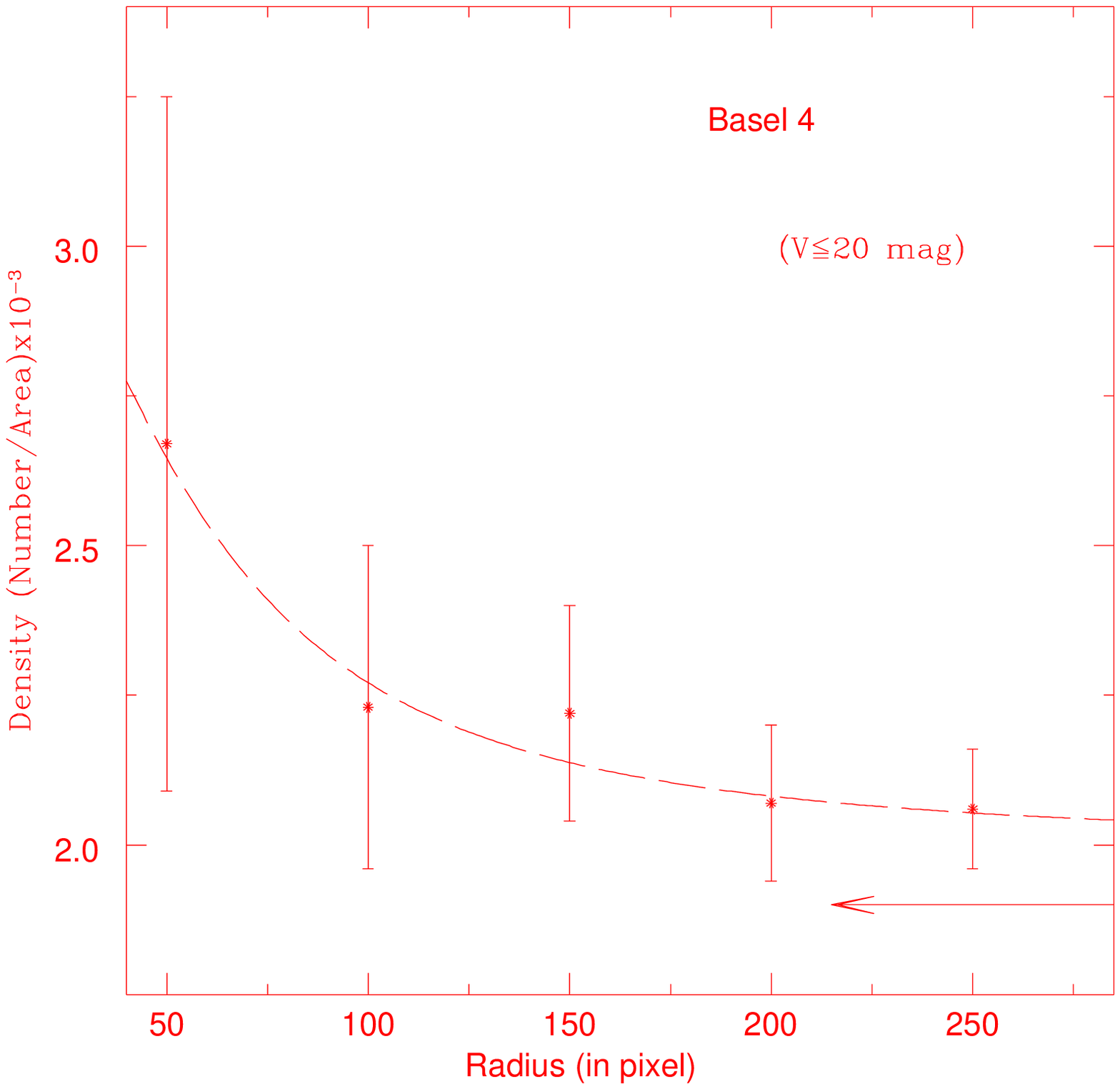 ,width=9.5cm,height=8.5cm}
\psfig{figure=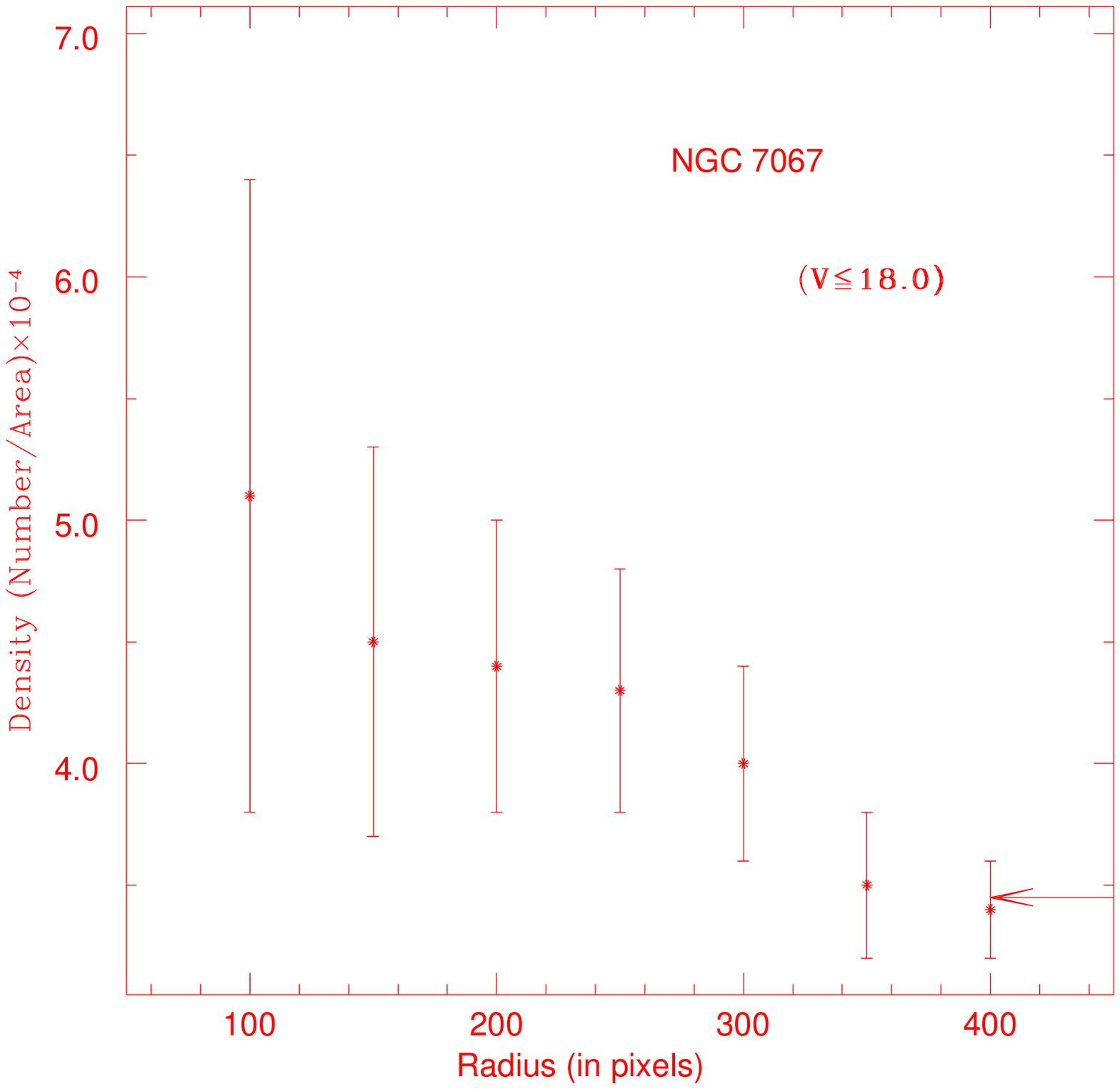 ,width=9.5cm,height=8.5cm}
\caption{Radial density profile for Basel 4 and NGC 7067. The length
of the errorbar denotes errors resulting from sampling statistics 
(=$\frac{1}{\sqrt{N}}$ where N is the number of stars used in the density 
estimation at that point). Dotted curves represent fitted profile and arrows 
represent the level of field star densities.}
\end{figure}

For a reliable determination of cluster parameters it is essential to know about the 
radial extent of the cluster. According to Mermilliod (1995), Basel 4 and NGC 7067 
has a diameter of 5 and 3 arcmin, so our study covers the entire cluster region. 
To estimate the cluster radius, we derive the surface stellar density by 
performing star counts in concentric rings around the estimated center of the 
cluster, and then dividing by their respective areas. The center of the cluster 
is determined iteratively by calculating average X and Y position of the stars 
within 300 pixels from an eye estimated center, until they converged to a 
constant value. The pixel coordinate of the cluster center obtained in this 
way are (600, 780) and (690, 700) for Basel 4 and NGC 7067 respectively.
 There may be few tens of pixels error in locating the cluster center. Center of 
the clusters are marked by "C" in the identification maps (Fig 1). The radial density profile and the 
corresponding Poisson error bars are depicted in Fig 3. A clear radius density 
gradient present in Basel 4 confirms the existence of clustering. In NGC 7067 
the density profile decreases with radius slowly up to the limit of the covered region. Following 
Kaluzny (1992), we describe the $\rho$(r) of an open cluster as:
\begin{displaymath}
~~~~~~~~~~~~~~~~~~~ \rho(r) \propto \frac {f_0}{1+(r/r_{c})^2},
\end{displaymath}
where the cluster core radius $r_{c}$ is the radial distance at which the
value of $\rho(r)$ becomes half of the central density ${f_0}$. We fit this
function to the observed data points in the cluster Basel 4 and use $\chi^{2}$ 
minimization technique to determine $r_{c}$ and other constants. As can be 
seen in Fig 3, the fitting of the function is satisfactory for Basel 4. 
Such fitting could not be done in NGC 7067 due to large errors in the values of 
$\rho(r)$. In the cluster Basel 4, values of $\rho(r)$ flatter 
at larger radii indicating probably the level of field star 
density in the cluster direction shown by arrow in Fig 3. The field star density 
thus obtained is $2.1\times10^{-3}$ per pixel$^{2}$ for Basel 4. In the case of NGC 7067, 
the field star density has been estimated as $0.3\times10^{-3}$ per pixel$^{2}$ based on 
the last two data points of its radial density profile in Fig 3. The radius at which 
the value of $\rho$ 
becomes approximately equal to the field star density has been considered as the
cluster radius. The values determined in this way are 1$^\prime$.8 and 3$^\prime$.0 
for Basel 4 and NGC 7067 respectively. Present radius estimates is lower  
with the value listed in Table 1 for Basel 4 while in the case of NGC 7067 our 
estimated value is larger (see Table 1).

As the observed area is much larger than the cluster area, we have considered stars 
having more than 2.5 and 1.6 cluster radius as field stars for the cluster Basel 4 
and NGC 7067 respectively (see Fig 1). The areas of the field 
regions are 4.2$\times$10$^5$ and 3.1$\times$10$^5$ pixel$^2$ for the cluster Basel 4 
and NGC 7067 respectively. For Basel 4, the closest boundary of the field region is 
about 4$^{\prime}$.5 away from the cluster center in the west direction while for NGC 
7067, it is about 5$^{\prime}$.0 away in the north direction. For the further analysis 
we have considered the stars laying within cluster radius.

\begin{figure*}
\hspace{0.0cm}\psfig{file=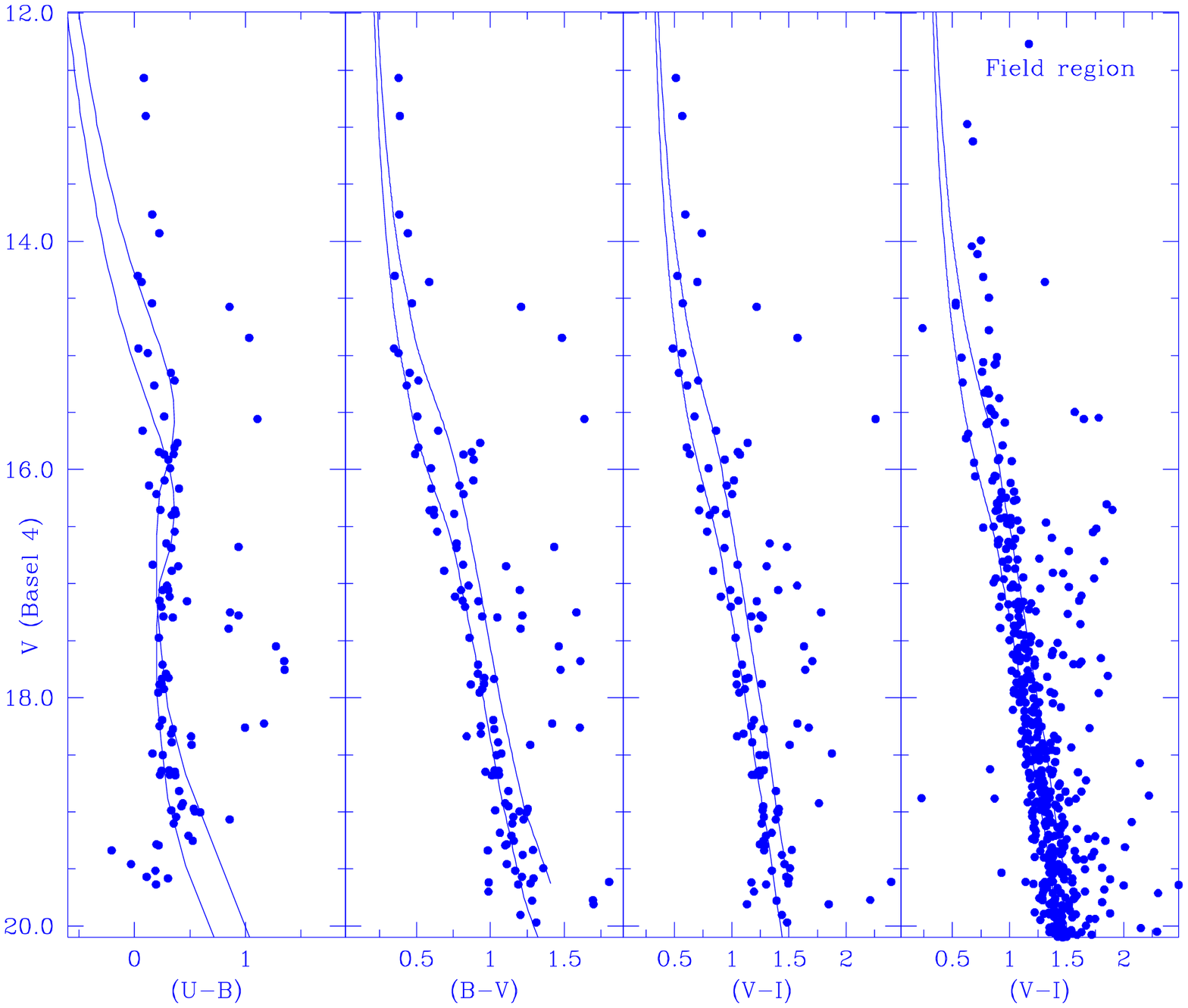,width=12cm,height=10.0cm}
\vspace{-1.0cm}
\hspace{3.0cm}\psfig{file=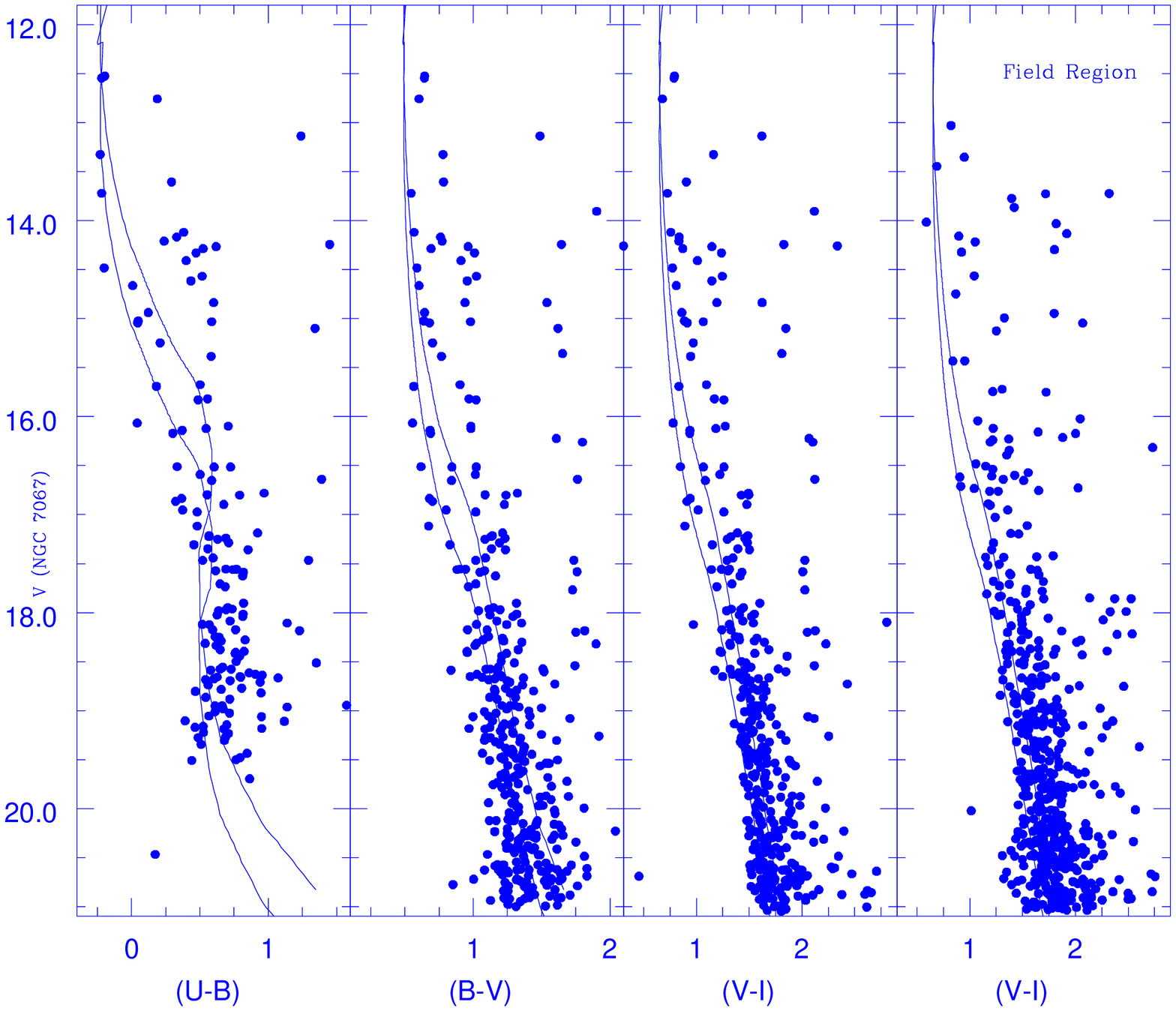,width=12cm,height=10.0cm}
\caption{The $V$, $(U-B)$; $V$, $(B-V)$ and $V$, $(V-I)$
diagrams for the stars observed by us in Basel 4 and NGC 7067 cluster regions and $V$, $(V-I)$ 
CM diagrams of the corresponding field regions. Solid lines 
represent the blue and red envelope of the cluster MS. The red envelope is determined by shifting 
blue envelope vertically with 0.80 mag. }
\label{fig3.7}
\end{figure*}

\begin{table*}
\centering
\begin{minipage}{140mm}
\caption{Frequency distribution of the stars in the $V$, $(V-I)$
diagram of the cluster and field regions. $N_{B}$, $N_{S}$ and $N_{R}$ denote
the number of stars in a magnitude bin blueward, along and redward of the
cluster sequence respectively. The number of stars in the field regions are
corrected for area differences. $N_{C}$ (difference between the $N_{S}$ value
of cluster and field regions) denotes the statistically expected number of
cluster members in the magnitude bin.}
\vspace{0.7cm}
\begin{tabular}{|c|ccc|ccc|c|ccc|ccc|c|}
\hline
&\multicolumn{7}{|c|}{Basel 4} & \multicolumn{7}{c|}{NGC 7067} \\
\cline{2-15}
$V$ range &\multicolumn{3}{|c|}{Cluster region} & \multicolumn{3}{|c|}{Field
region}& &
\multicolumn{3}{|c|}{Cluster region} & \multicolumn{3}{|c|}{Field region}&  \\
&$N_{B}$&$N_{S}$&$N_{R}$&$N_{B}$&$N_{S}$&$N_{R}$&$N_{C}$&$N_{B}$&$N_{S}$&$N_{R}$
&$N_{B}$&$N_{S}$ &$N_{R}$ &$N_{C}$ \\
\hline
12 - 13 & 0 & 2& 0 & 0 & 0 & 0 & 2 & 0  & 1 &2 & 0& 0 & 0  & 1 \\
13 - 14 & 0 & 2& 0 & 0 & 0 & 0 & 2 & 0  & 2 &3 & 1& 1 & 4  & 1 \\
14 - 15 & 0 & 4& 3 & 0 & 0 & 1 & 4 & 0  & 4 &11& 0& 0 & 6  & 4  \\
15 - 16 & 0 & 8& 5 & 0 & 2 & 3 & 6 & 0  & 3 &8 & 0& 1 & 4  & 2 \\
16 - 17 & 0 &11& 4 & 1 & 5 & 3 & 6 & 0  &12 &8 & 0& 4 & 16 & 8 \\
17 - 18 & 0 &16& 8 & 4 & 7 & 4 & 9 & 0  & 17&11&1 & 7 & 19 & 10 \\
18 - 19 & 2 &17& 5 & 7 & 9 & 2 & 8 & 5  & 35&20&2 &17 & 36 & 18\\
19 - 20 & 3 &22& 3 & 9 & 9 & 1 & 13& 18 & 44&19&10&23 & 61 & 21 \\
\hline
\end{tabular}
\end{minipage}
\end{table*}

\begin{figure}
\psfig{figure=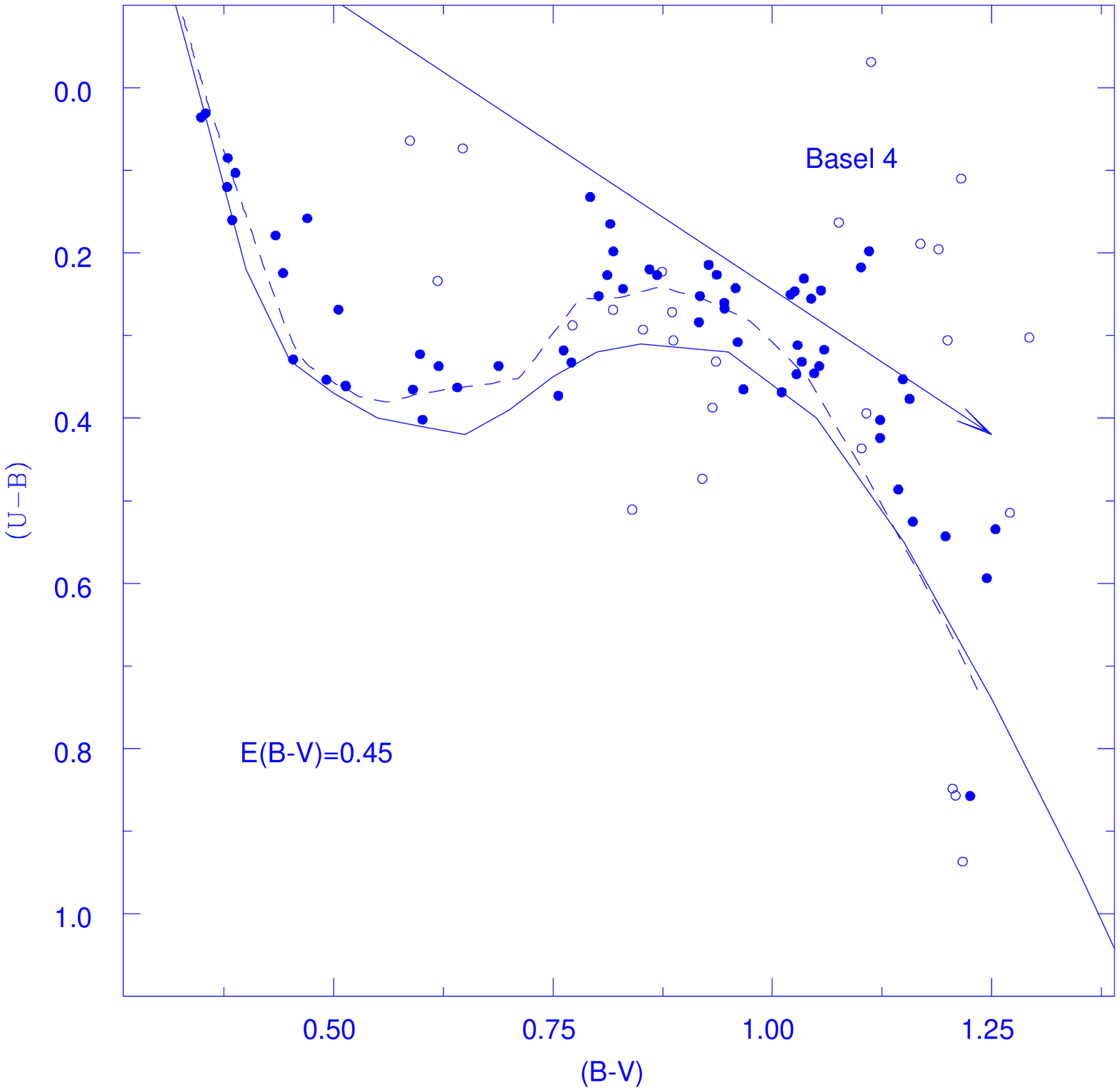,height=8.0cm,width=8.0cm}
\psfig{figure=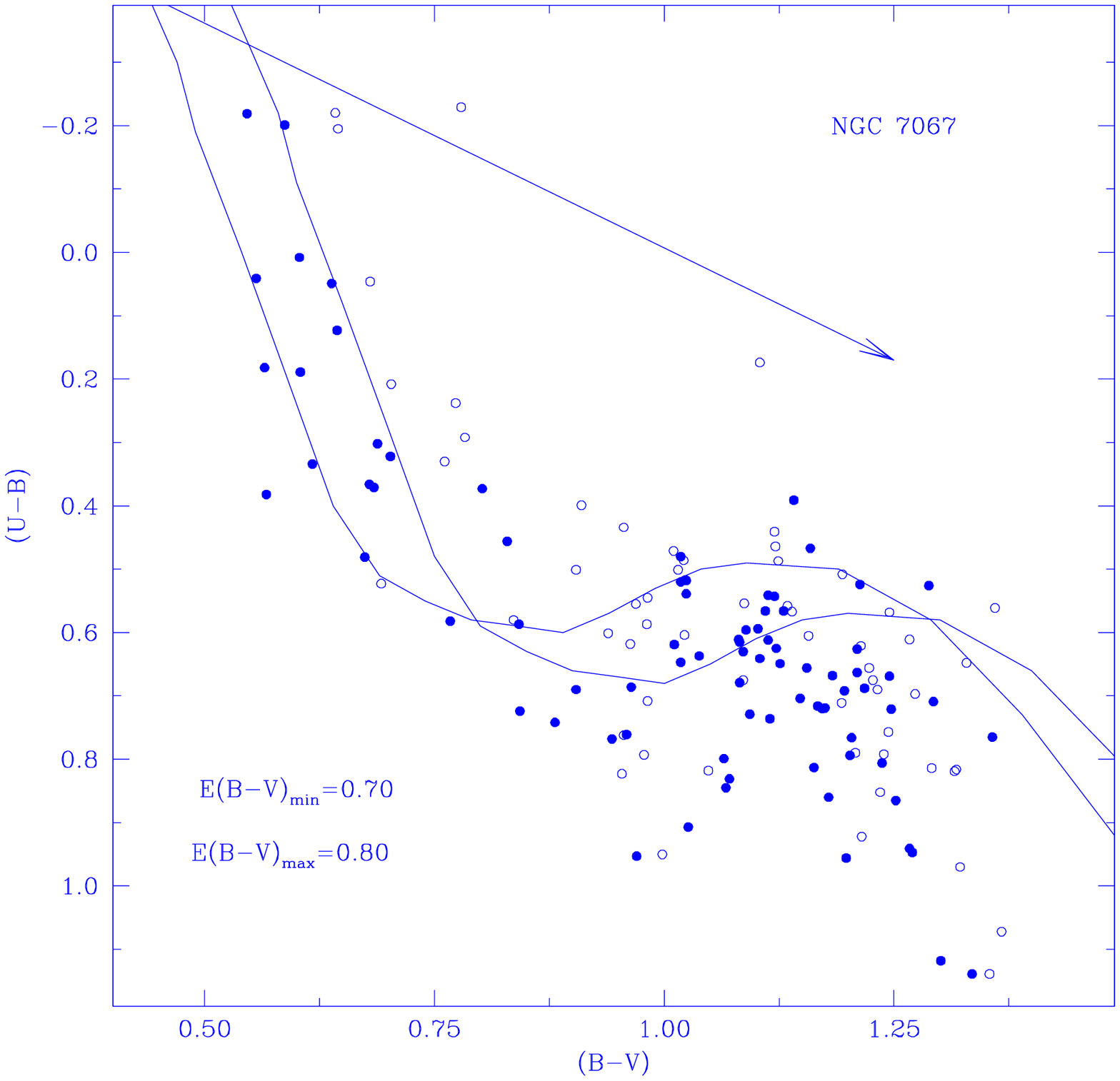,height=8.0cm,width=8.0cm}
\caption{The $(U-B)$ versus $(B-V)$ diagrams of the stars in cluster region
for Basel 4 and NGC 7067. Stars considered as non member in Fig. 4 are shown as 
open circles. The arrow represents slope 0.72
and direction of the reddening vector. The solid curve represents the
locus of Schmidt - Kaler's (1982) ZAMS fitted for the marked values of colour
excesses. The curve shown by short dash lines in Bessel 4 is the ZAMS given by 
Schaerer et al. (1993) for Z = 0.008}
\end{figure}

\begin{figure}
\psfig{file=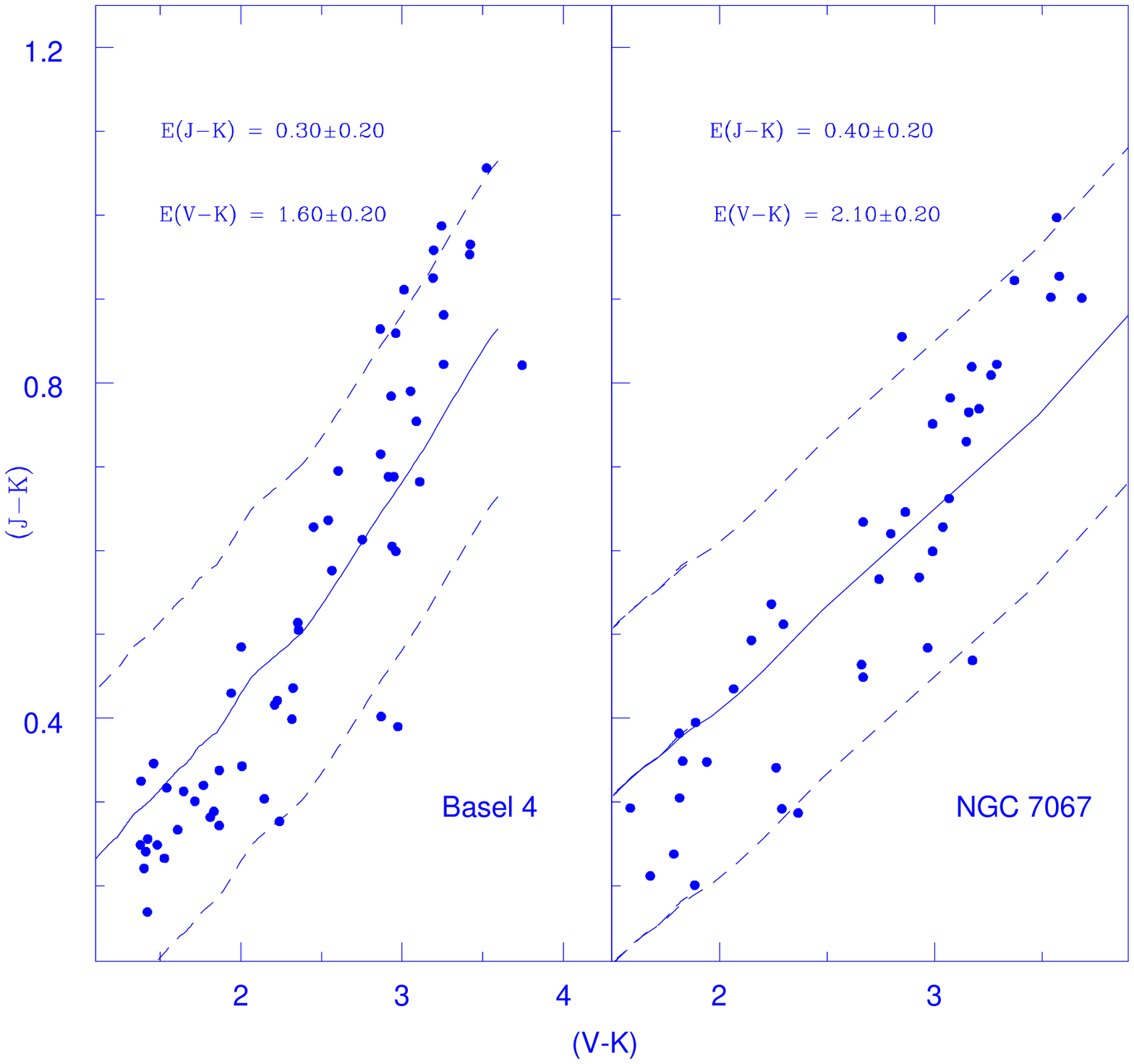,width=8.5cm,height=7cm}
\caption{The $(J-K)$ versus $(V-K)$ colour - colour diagram of all the stars
 which are common in $V$ and $JHK$ data within the cluster radius for the 
cluster Basel 4 and NGC 7067. The solid line is a ZAMS fitted for the marked 
values of colour excesses. Dashed lines show the errorbar.}
\end{figure}

\subsection {Apparent CM diagrams of the cluster regions}

The apparent CM diagrams of Basel 4 and NGC 7067 for the stars present within the 
cluster radius are shown in Fig 4. The detection in $U$ filter is 
not as deep as in $BVRI$ because of the low quantum efficiency of the CCD detector 
in $U$ region. So, there are large number of stars without any $U$ measurements. A 
well defined cluster MS contaminated by field stars
 is clearly visible in the CM diagrams of Basel 4 while in NGC 7067 the MS is not 
so well populated because of the poorness of the cluster. In the cluster NGC 7067, 
the stars on the red side of the MS appear to form a sequence parallel to the MS 
and that can be ascribed due to Galactic disk field stars. The field star contamination 
increases with decreasing brightness. The cluster sequence fainter than $V$ $\sim$ 16 mag
in Basel 4 and $V$ $\sim$ 18 mag in NGC 7067 have larger scatter. This may be due 
to photometric errors as well as field star contamination. It is difficult to 
separate field stars from the cluster members. For the separation of cluster members from
the field stars, precise proper motion and/or radial velocity measurements of
these stars are required. In the absence of such data for these clusters, we selected 
members by defining the binary sequence. It has been defined by shifting the blue envelope 
with 0.8 mag vertically which is shown in the CM diagrams of the clusters. Some stars, in 
spite of their ambiguous positions could not definitively be rejected as likely cluster members. 
From the $V$, $(V-I)$ diagram of the field region, statistically expected number of field stars 
among the photometric cluster members has been given in Table 6. The frequency distribution of 
the field star contamination in different part of the CM diagram can be estimated from the Table 6. 
It is thus clear that all photometric probable members can not be cluster members and non-members 
should be subtracted in the studies of cluster MF etc. However, probable members located within a 
cluster radius from its center can be used to determine the cluster parameters, as they have 
relatively less field star contamination and this has been done in the sections to follow.

\subsection {Interstellar extinction towards the clusters}
Fig 5 shows the $(U-B)$ versus $(B-V)$ diagrams for determining the interstellar 
extinction using the probable cluster members. We fit the intrinsic zero-age main-sequence 
(ZAMS) given by Schmidt-Kaler (1982) valid for stars of luminosity class V to the MS stars 
of spectral type earlier than A0 assuming the slope of reddening $E(U-B)/E(B-V)$ as 0.72. 

In the cluster Basel 4, ZAMS given by Schmidt-Kaler (1982) is not fitting well for the stars 
of spectral type A, F and G. Excess in $(U-B)$ colour is clearly visible for 
the stars of $(B-V) > 0.50$ mag. This indicates that the cluster is metal deficient. 
 The UV excess $\delta(U-B)$ determined with respect to Hyades MS turns out to be $\sim$ 0.1 mag. 
Using the [Fe/H] versus $\delta(U-B)$ relation of Carney (1979) we estimated [Fe/H] $\sim -0.35$ 
which correspond to Z $\sim$ 0.008. To estimating the reddening in the direction of this cluster 
we therefore fitted the ZAMS given by Schaerer et al. (1993) for Z $=$ 0.008 which is shown by 
short dash lines in the two colour diagram of Basel 4. The ZAMS of Z $=$ 0.008 fits nicely and 
provide the reddening $E(B-V) = 0.45\pm0.05$ for this cluster which is in agreement with the 
earlier findings (see Table 1).

Unlike Basel 4, in the cluster NGC 7067, ZAMS given by Schmidt-Kaler (1982) for the solar
metallicity fits both early and late type stars. The fitted values of $E(B-V)$ vary from 
0.70 to 0.80 mag. The mean value is $E(B-V)= 0.75\pm$0.05 mag. Our mean reddening estimate 
for the imaged region agree fairly well with values estimated earlier by others (see Table 1). 

We investigate the nature of interstellar extinction law towards the clusters, 
by considering the stars having spectral type earlier than A0. This has
been selected from their position in the $(U-B)$ versus $(B-V)$ and apparent 
CM diagrams which reveals that bright stars with $V$$<$ 16.0 mag and
$(B-V)$$<$0.60 mag in Basel 4 and with $V$$<$ 16.5 mag and $(B-V)<0.75$ mag in NGC 7067 are needed 
stars. The number of such stars are 11 and 12 in Basel 4 and NGC 7067 respectively. 
The intrinsic colours for these stars have been determined using 
$UBV$ photometric Q-method (cf. Johnson \& Morgan 1953; Sagar \& Joshi 1979) 
and the calibrations given by Caldwell et al. (1993)
for $(U-B)_{0}$, $(V-R)_{0}$ and $(V-I)_{0}$ with $(B-V)_{0}$. The mean values
of the colour excess ratios derived in this way are
listed in Table 7 for both the clusters. They indicate that the law of interstellar 
extinction in the direction of the clusters under discussion is normal.

\begin{table}
 \centering
 \caption{A comparison of the colour excess ratios with $E(B-V)$ for both star
clusters with the corresponding values for the normal interstellar extinction
law given by Cardelli et al. (1989).}

\begin{tabular}{cccc}
\hline
Objects&$\frac{E(U-B)}{E(B-V)}$&$\frac{E(V-R)}{E(B-V)}$&$\frac{E(V-I)}{E(B-V)}$\\
\hline
Normal interstellar&0.72&0.65&1.25\\
Basel 4&0.71$\pm$0.05&0.68$\pm$0.04&1.33$\pm$0.10\\
NGC 7067&0.69$\pm$0.04&0.56$\pm$0.02&1.33$\pm$0.08\\
\hline
\end{tabular}
\end{table}

\begin{figure}
\psfig{file=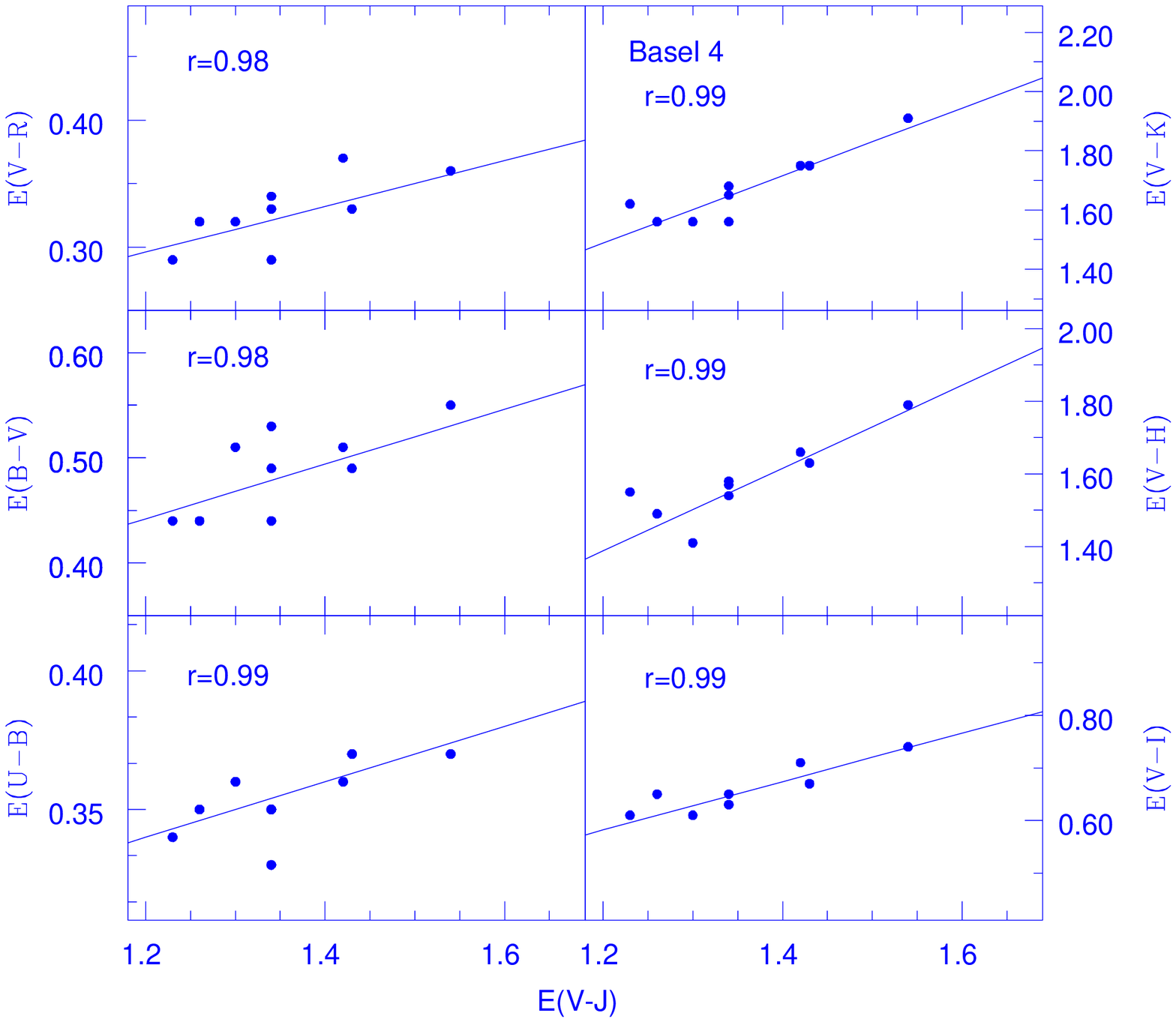,width=8.5cm,height=10cm}
\psfig{file=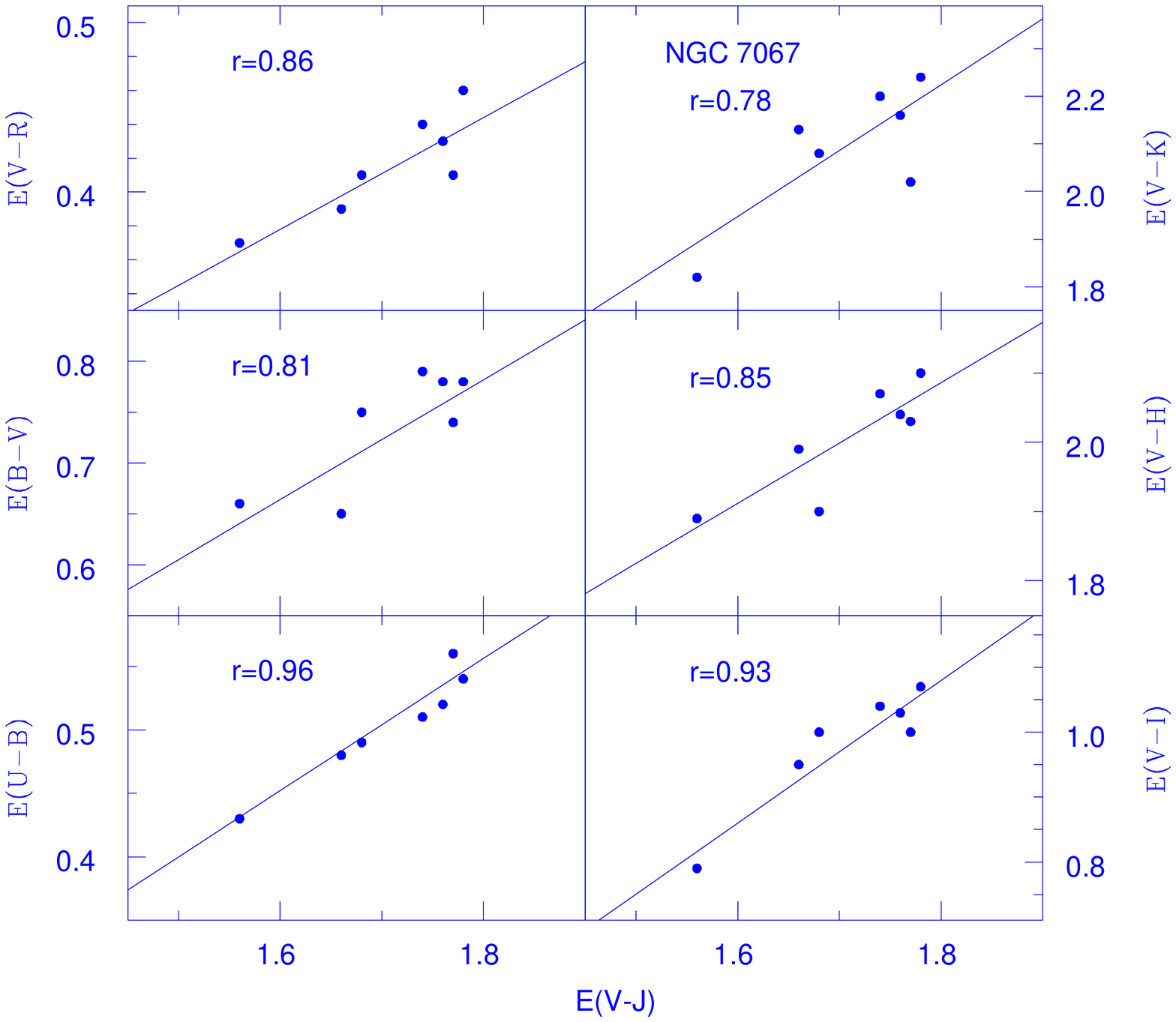,width=8.5cm,height=10cm}
\caption{The plot of $E(U-B)$, $E(B-V)$, $E(V-R)$ $E(V-I)$, $E(V-H)$ and
$(V-K)$ against $E(V-J)$ for Basel 4 and NGC 7067. Solid line in each diagram
represents least square linear fit to the data points. The values of correlation
 coefficients are shown in the diagram.}
\end{figure}

\subsubsection {Interstellar extinction in near - IR}

By using the optical and infrared data, we estimated the interstellar extinction for 
both clusters under study. There are 65 and 44 common stars in the cluster Basel 4 
and NGC 7067 within the cluster radius respectively. Fig. 6 shows the $(J-K)$ vs
$(V-K)$ diagrams and fit a ZAMS for metallicity Z = 0.008 taken from Schaerer et al. 
(1993) in the cluster Basel 4 and Z = 0.02 taken from Schaller et al. (1992) in the 
cluster NGC 7067. This gives $E(J-K)$ = 0.30$\pm$0.20 mag and $E(V-K)$ = 
1.60$\pm$0.20 mag for the cluster Basel 4 and $E(J-K)$ = 0.40$\pm$0.20 mag and
$E(V-K)$ = 2.10$\pm$0.20 mag for the cluster NGC 7067. For both clusters
the ratio $\frac{E(J-K)}{E(V-K)}$ $\sim$ 0.20$\pm$0.30 is in good agreement with 
the normal interstellar extinction value 0.19 suggested by Cardelli et al. (1989). 
However, scattering is larger due to the error size in $JHK$ data.

\begin{figure}
\hbox{
\hspace{-1.4cm}\psfig{file=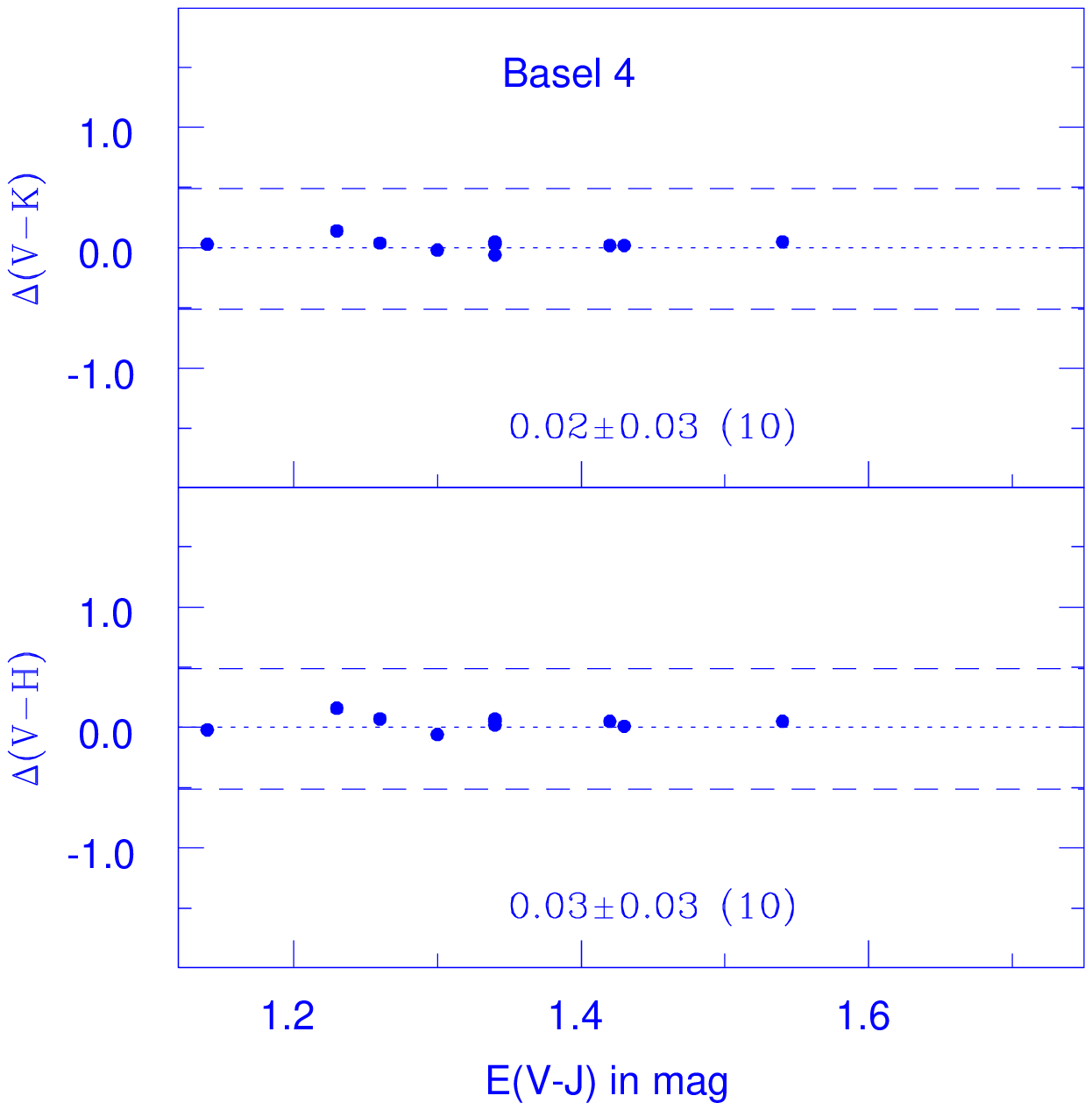 ,width=7.0cm,height=9.0cm}
\hspace{-2.5cm}\psfig{file=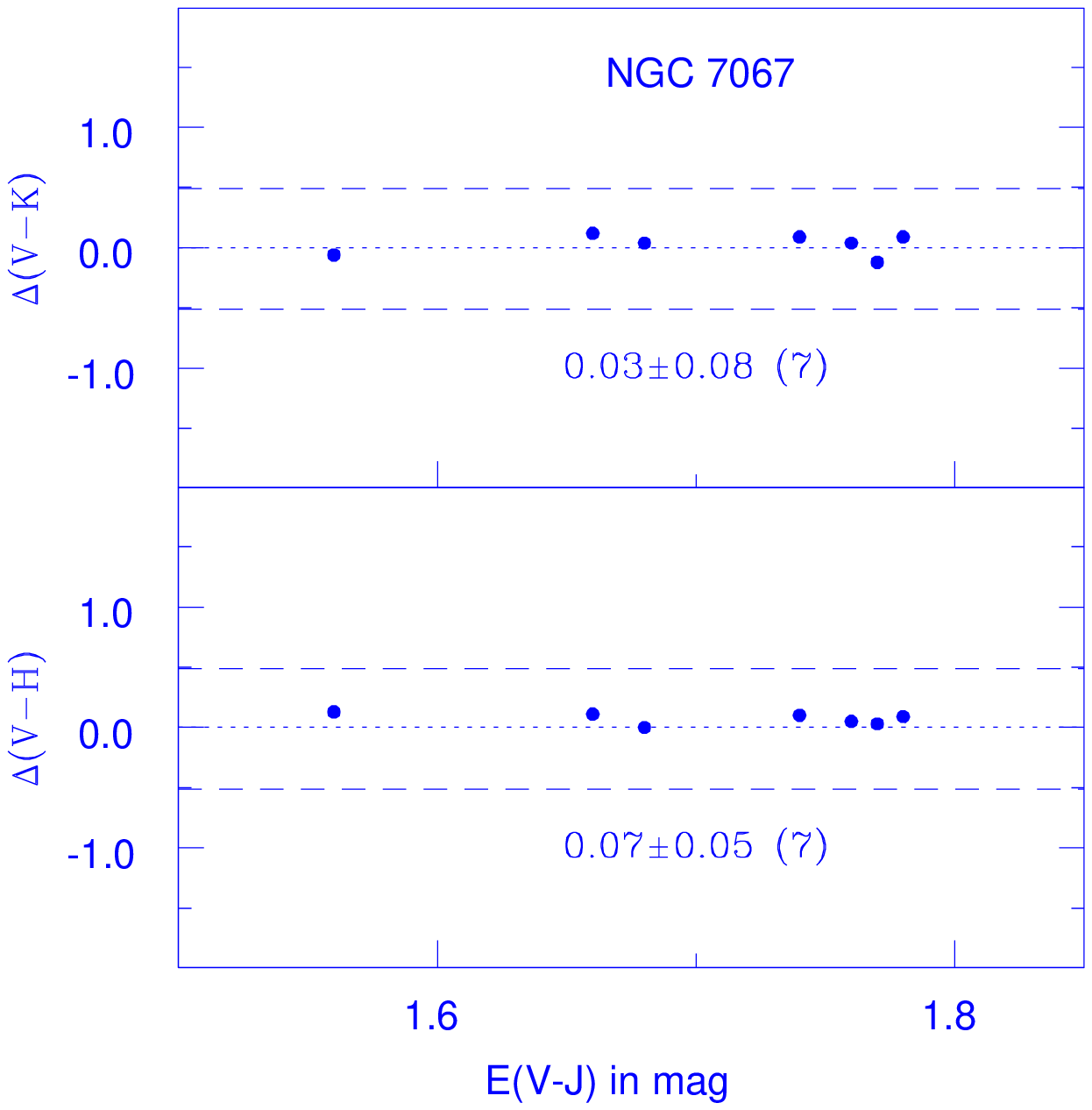 ,width=7.0cm,height=9.0cm}
}
\vspace{-1.5cm}
\caption{Plots of near-IR flux excess/deficiency in terms of $\Delta(V-H)$ and 
$\Delta(V-K)$ against the colour excess $E(V-J)$ for Basel 4 and NGC 7067. The horizontal 
dotted lines denote zero excess. The short dashed lines denote the extent of the expected 
errors.}
\end{figure}

\subsubsection {The law of interstellar extinction}

We used the colour excess ratio method described by Johnson (1968) for the study of 
interstellar extinction law in the direction of both clusters under study. 
For this we used the stars of spectral type earlier than A0. We determined 
the colour excesses by comparing the observed colours of the stars with its intrinsic 
colours derived from the MKK spectral type-luminosity class colour relation given by 
FitzGerald (1970) for $(U-V)$ and $(B-V)$; by Johnson (1966) for $(V-R)$ and $(V-I)$; and by Koornneef
(1983) for $(V-J)$, $(V-H)$ and $(V-K)$. For normalisation, we selected the 
$E(V-J)$ colour excess due to reasons described in Yadav \& Sagar (2002). In Fig. 7 we plot the
colour excess $E(U-B)$, $E(B-V)$, $E(V-R)$, $E(V-I)$, $E(V-H)$ and $E(V-K)$ against
$E(V-J)$. In this Fig, straight line represents the least square linear fits to the
data points. The values of correlation coefficient (r) and fit indicate that 
the data points are well represented by linear relation. The slopes of these 
straight lines as given in Table 8 represent reddening directions in the form of 
colour excess ratios. For comparison, the colour excess ratios given by Cardelli 
et al. (1989) for the normal interstellar matter are also listed in the Table 8. 
The present reddening directions 
agree well with those.

In addition to this we have also estimated the value of R to know about the nature 
of interstellar extinction law in the direction of clusters under study. We used 
the relation R = 1.1$E(V-K)$/$E(B-V)$ (Whittet \& Breda 1980) which is generally
used at longer wavelengths. The average values of R = 3.51
$\pm$0.30 (sd) and 3.04$\pm$0.22 (sd) for Basel 4 and NGC 7067 respectively are not too 
different from the value 3.1 for normal extinction law.\\ 

In the light of above analysis, we conclude that  interstellar extinction law is normal towards
both Basel 4 and NGC 7067 in agreement with our earlier result.

\subsubsection {Near-IR excess Fluxes}
An infrared excess is produced by the stars which are having their own envelope of gas 
and dust. To investigate the near-IR flux in the stars of the clusters under study, 
we plotted 
$\Delta(V-H)$ and $\Delta(V-K)$ against $E(V-J)$ in Fig 8. The differences 
between the observed colour excess in $(V-H)$ and $(V-K)$ based on spectral 
classification and the derived colour excess from $E(V-J)$ assuming normal 
extinction law are calculated. The differences can be considered statistically 
significant only if their absolute values are larger than $\sim$0.5 mag. The short 
dashed lines in Fig 8 represent the extent of expected errors. Observational 
uncertainties in $JHK$ magnitudes, inaccuracies in estimation of $E(V-J)$ and 
errors in spectral classification may play major role in the determination of 
differences. An inspection of Fig 8 leads that the absolute values of 
$\Delta(V-H)$ and $\Delta(V-K)$ are close to zero of all the members. 
This indicate that there is no signature of near-IR excess fluxes.

\begin{figure*}
\hspace{0.0cm}\psfig{figure=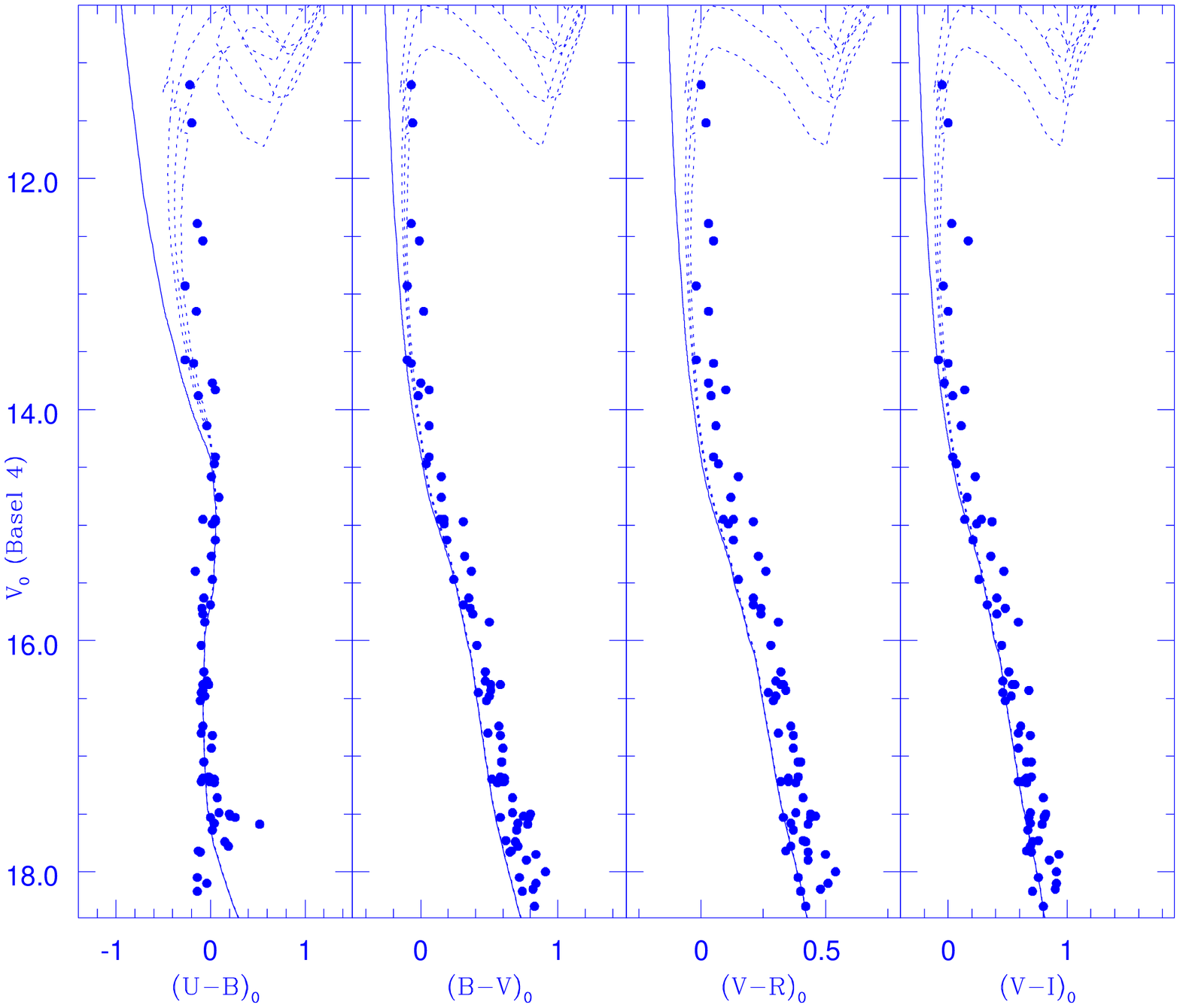,height=8cm,width=12cm}
\hspace{2.0cm}\psfig{figure=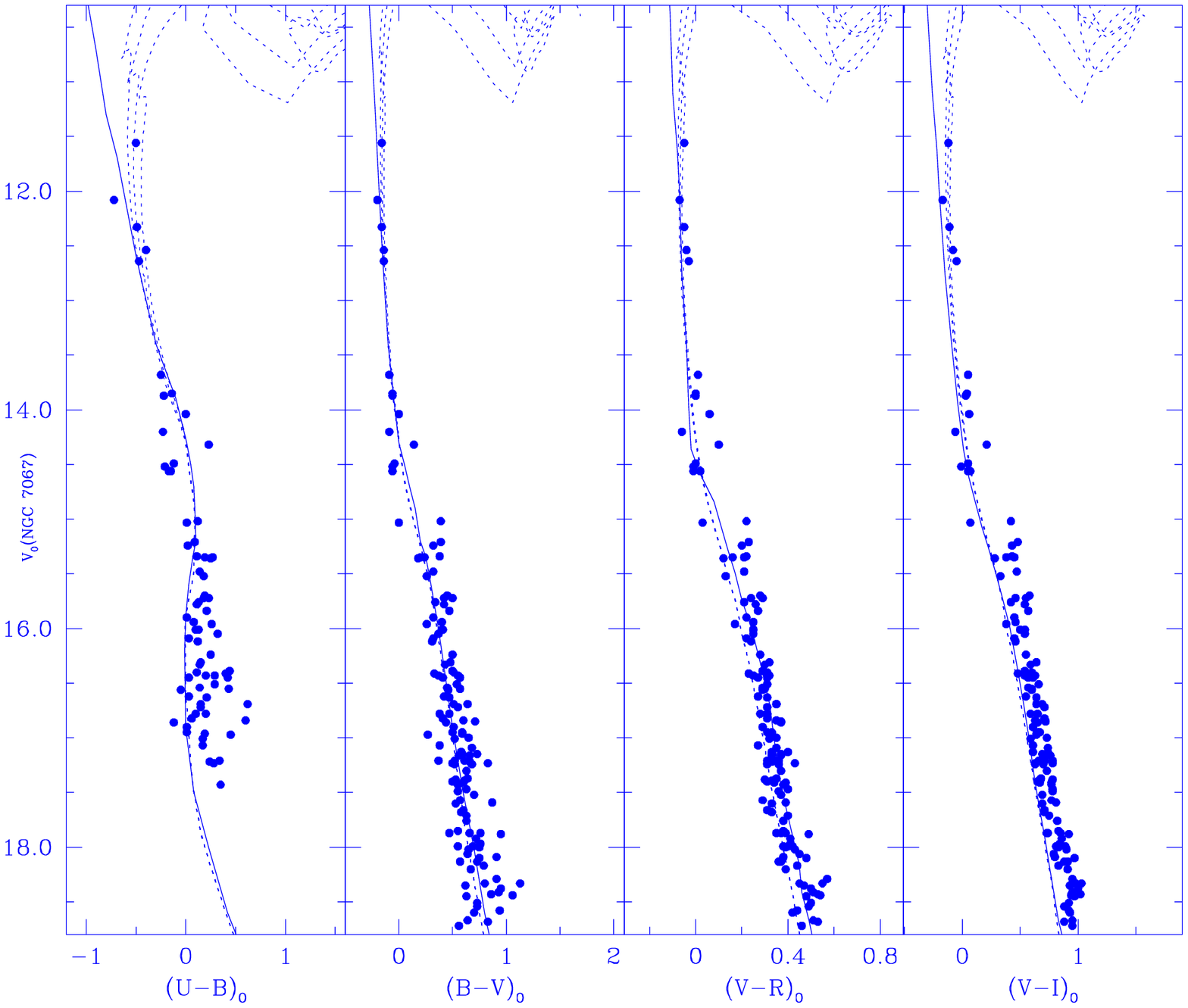,height=8cm,width=12cm}
\caption {The $V_{0}$, $(U-B)_{0}$; $V_{0}$, $(B-V)_{0}$; $V_{0}$, $(V-R)_{0}$ and $V_{0}$,
$(V-I)_{0}$ diagrams for stars of the Basel 4 and NGC 7067. The continuous curves, ZAMS fitted
to the MS and dotted curves are the isochrones for Z $=$ 0.008 stars of log(age)=8.2, 8.3 and 8.4 for 
Basel 4 and Z $=$ 0.02 stars of log(age)=7.9, 8.0 and 8.1 for NGC 7067.}
\end{figure*}

\subsection {Distance to the clusters}

The ZAMS fitting procedure was employed to derive the distances of the clusters. Fig 9 
shows the intrinsic CM diagrams for Basel 4 and NGC 7067 which is plotted by considering  
the probable cluster members. For converting apparent $V$ magnitude and $(U-B)$, 
$(B-V)$, $(V-R)$ and $(V-I)$ colours into intrinsic one, we used average values of 
$E(B-V)$ and following relations for $E(U-B)$ (cf. Kamp 1974; Sagar \& Joshi 1979), 
A$_{v}$ and $E(V-I)$ (Walker 1987) and $E(V-R)$ (Alcal\'{a} et al. 1988).\\

$E(U-B)$ = [X + 0.05$E(B-V)$]$E(B-V)$

\vspace{0.3cm}
where X = 0.62 $-$ 0.3$(B-V)$$_{0}$ for $(B-V)$$_{0}$ $<$ $-$0.09

~~and~~~X = 0.66 + 0.08$(B-V)$$_{0}$ for$(B-V)$$_{0}$ $>$ $-0.09$\\

          A$_{v}$ = [3.06 + 0.25$(B-V)$$_{0}$ + 0.05$E(B-V)$]$E(B-V)$;\\
and      $E(V-R)$ = [E1 + E2E$(B-V)$]$E(B-V)$\\

where E1 = 0.6316 + 0.0713$(B-V)$$_{0}$\\
and E2 = 0.0362 + 0.0078$(B-V)$$_{0}$;\\

$E(V-I)$ = 1.25[1 + 0.06$(B-V)$$_{0}$ + 0.014$E(B-V)$]$E(B-V)$\\

The ZAMS for Z $=$ 0.008 is plotted in $V_{0}$, $(U-B)$$_{0}$; $V_{0}$, $(B-V)$$_{0}$; $V_{0}$, 
$(V-R)$$_{0}$ and $V_{0}$, $(V-I)$$_{0}$ diagrams of Basel 4 are taken from Schaerer et al. (1993). 
In $V_{0}$, $(U-B)$$_{0}$ and $V_{0}$, $(B-V)$$_{0}$ diagrams of NGC 7067, we fitted the solar
metallicity ZAMS given by Schmidt-Kaler (1982) while that given by Walker (1985)
was fitted in $V_{0}$, $(V-I)$$_{0}$ diagram. For $V_{0}$, $(V-R)$$_{0}$ 
diagram, we have calculated $(V-R)$$_{0}$ using its relation with $(B-V)_{0}$ 
given by Caldwell et al. (1993). The good fitting of the ZAMS to the intrinsic CM 
diagrams was achieved for a distance modulus $(m-M)$$_{0}$ as 12.4$\pm$0.2 and 
12.8$\pm$0.2 mag for Basel 4 and NGC 7067 respectively. The corresponding distances 
are 3.0$\pm$0.2 Kpc and 3.6$\pm$0.2 Kpc. The fact that 
we were able to find faint probable cluster members allow us to get a better definition 
of the cluster lower main sequence which in turn improves the estimation of the 
distances. The distance of the cluster Basel 4 is estimated as 3.0$\pm$0.2 Kpc,
which is smaller than the value 5.6 Kpc given by Svolopoulos (1965) based on
RGU photographic photometry. We derived 3.6$\pm$0.2 Kpc distance for the cluster 
NGC 7067, which is also the value given by Lyng\aa~~(1987). The distance of this
cluster is 1.3 Kpc given in the catalogue of Dias et al. (2002),
which is much smaller than the value estimated by us.

\begin{table*}
\begin{minipage}{140mm}
\caption{A comparison of extinction law in the direction of Basel 4 and NGC 7067 with normal 
extinction law given by Cardelli et al. (1989).}
\begin{center}
\begin{tabular}{cccccccc}
\hline
Objects&$\frac{E(U-B)}{E(V-J)}$&$\frac{E(B-V)}{E(V-J)}$&$\frac{E(V-R)}{E(V-J)}$&$
\frac{E(V-I)}{E(V-J)}$&$\frac{E(V-H)}{E(V-J)}$&$\frac{E(V-K)}{E(V-J)}$&$\frac{E(
J-K)}{E(V-K)}$\\
\hline
Normal value&0.32&0.43&0.27&0.56&1.13&1.21&0.19\\
Basel 4&0.26$\pm$0.02&0.36$\pm$0.02&0.25$\pm$0.02&0.49$\pm$0.05&1.16
$\pm$0.04&1.24$\pm$0.04&0.20$\pm$0.30\\
NGC 7067&0.30$\pm$0.01&0.43$\pm$0.02&0.24$\pm$0.08&0.57$\pm$0.03&1.17
$\pm$0.03&1.22$\pm$0.05&0.20$\pm$0.30\\
\hline
\end{tabular}
\end{center}
\end{minipage}
\end{table*}

\subsection {Ages of the clusters}

The ages of the clusters are determined with the aid of the isochrones in the intrinsic 
CM diagrams (Fig 9) of Z = 0.008 taken from Schaerer et al. (1993) for Basel 4 and Z = 
0.02 taken from Schaller et al. (1992) for NGC 7067. The isochrones are computed by  
considering the mass loss and convective core 
overshooting. By fitting the isochrones to the cluster upper sequence of Fig 9, we 
found age of 200$\pm$50 and 100$\pm$25 Myr for Basel 4 and NGC 7067 respectively. 
 For the clusters Basel 4 and NGC 7067, our estimated values are larger than the values 
 given in Lyng\aa ~(1987) and Dias et al. (2002) catalogues.

Using the optical as well as near-IR data, we redetermined the distance and age of both 
the clusters. Fig 10 represents $V$ vs $(V-K)$ and $K$ vs $(J-K)$ CM diagrams. We have 
fitted the theoretical isochrones given by Schaerer et al. (1993) (Z = 0.008) of log(age) = 
8.3 for Basel 4 and by Schaller et al. (1992) (Z = 0.02) of log(age) = 8.0 for NGC 7067. The 
apparent distance moduli $(m - M)_{V, (V-K)}$ and 
$(m-M)_{K, (J-K)}$ turn out to be 13.9$\pm$0.3  and 12.5$\pm$0.3 mag 
for Basel 4 and 15.0$\pm$0.3 and 13.0$\pm$0.3 mag for NGC 7067 respectively. By using the 
reddening estimated in the previous section we derive a distance of 3.1$\pm$0.2 and 
3.7$\pm$0.2 Kpc for Basel 4 and NGC 7067 respectively. Both age and distance 
determination for both the clusters are thus in agreement with our earlier 
estimates. However, the scatter is larger because of the larger errors on the $JHK$ 
magnitudes.

\subsection {Luminosity and Mass function of the clusters}

Since, our first aim is to derive the luminosity function from star counts then 
it is necessary to draw our attention on the completeness of the derived star list.
Due to the stellar crowding on the CCD frame and efficiency of the data reduction 
programmes, not all the stars of the frame may be detected. For deriving the 
completeness factor (CF) we used $V$ vs $(V-I)$ CM diagram instead of others 
because it is deepest. Detail procedure of deriving CF is given below. 

\begin{figure}
\psfig{file=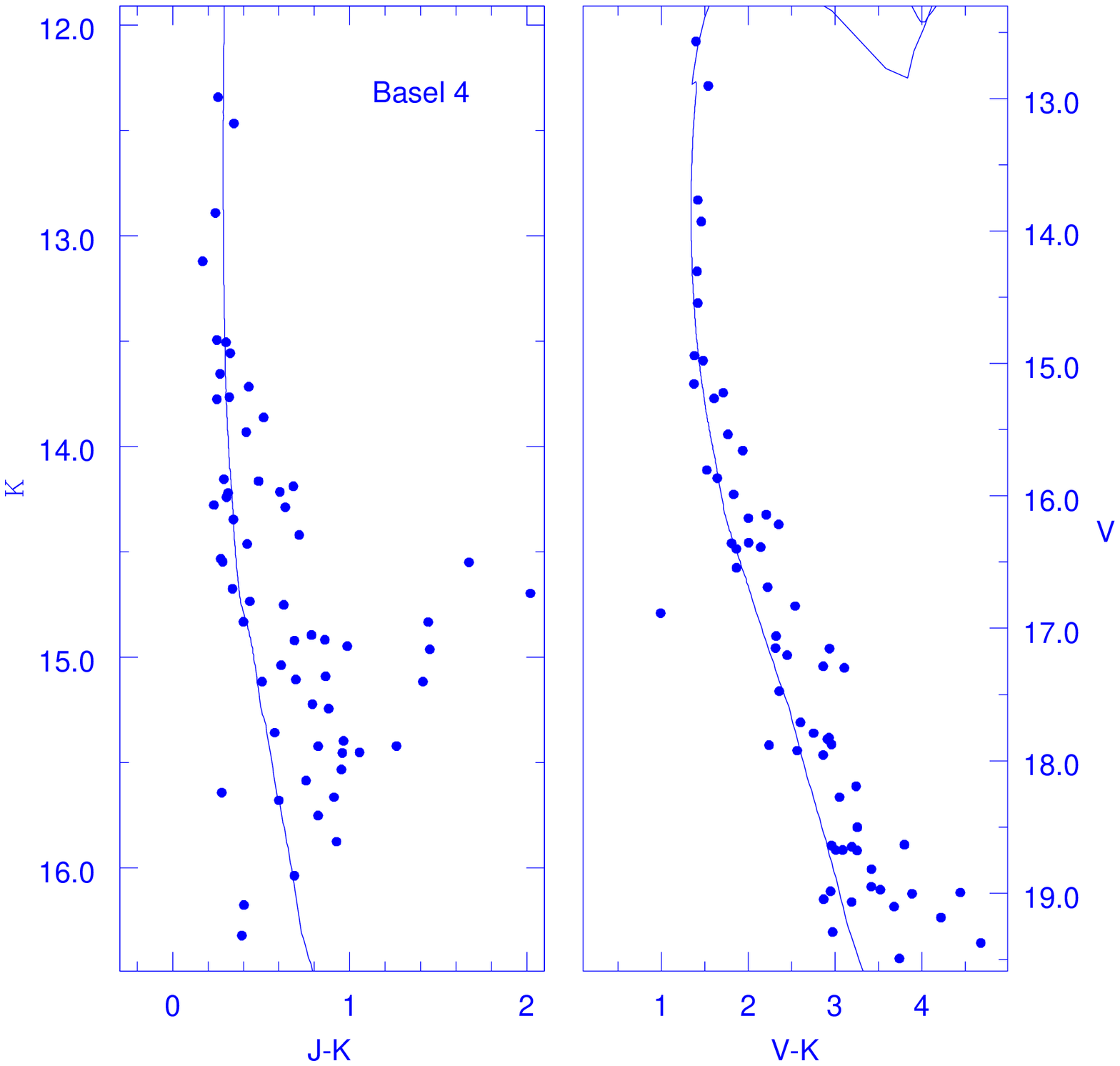 ,width=8.5cm,height=8cm}
\psfig{file=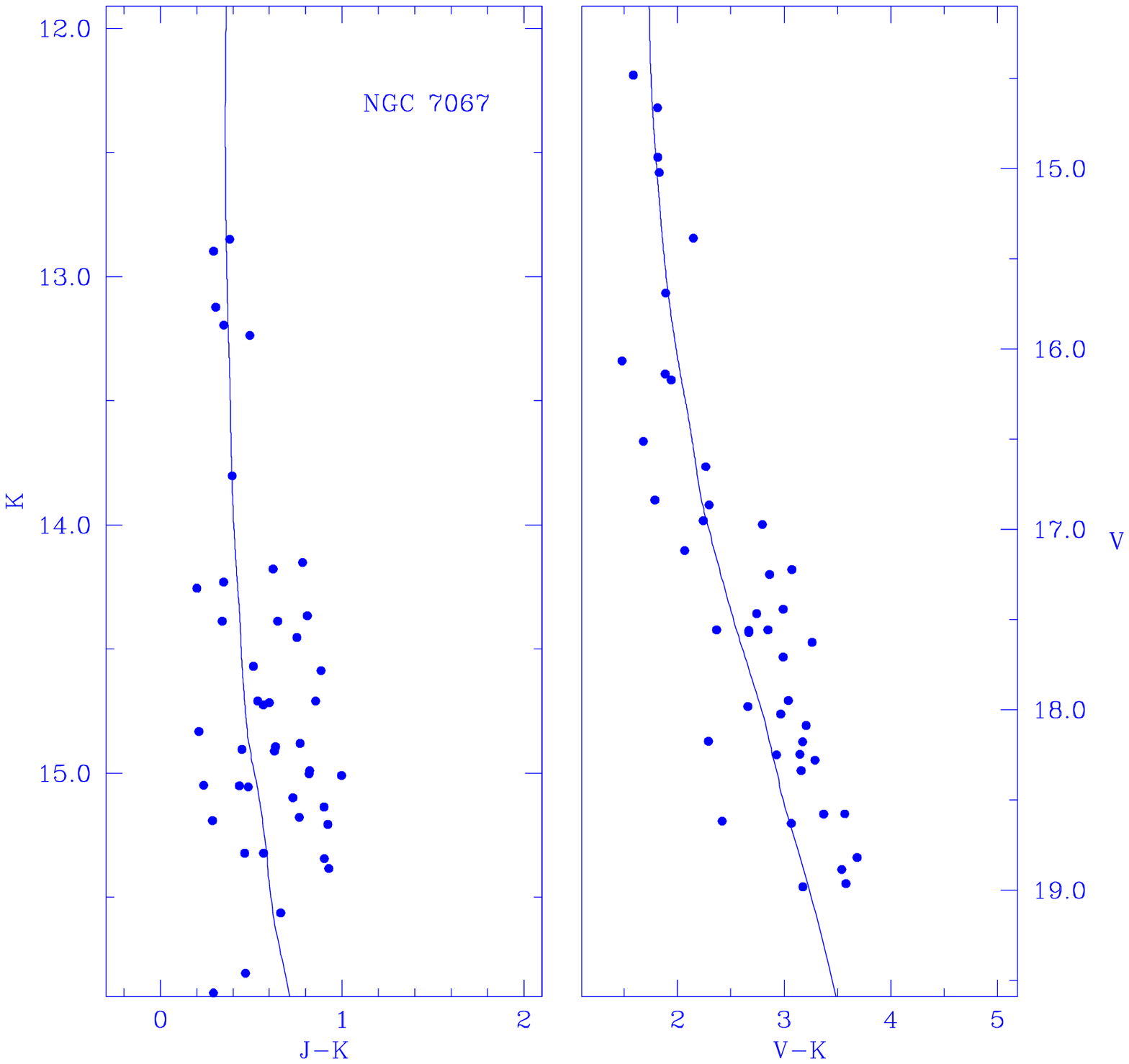 ,width=8.5cm,height=8cm}
\caption{The $K$, $(J-K)$ and $V$, $(V-K)$ diagrams of the sample stars in
the cluster Basel 4 and NGC 7067.}
\end{figure}

\subsubsection {Determination of photometric completeness}

To know about the completeness of our photometric data, we performed experiments 
with artificial stars using {\sl ADDSTAR} routine in {\sl DAOPHOT II}. In this
experiment a number of stars are added randomly in different
magnitude bin to the CCD original $V$ frame. For the $I$ band image
the added stars have same geometrical positions but differ in $I$
brightness according to mean $(V-I)$ colour of the MS stars. In order
to avoid overcrowding on the images by the
additional stars, we added only $10\%$ of the number of actually
detected stars. The luminosity distribution of the artificial stars
has been chosen in such a way that more stars are inserted into the
fainter magnitude bins. The photometric routines are run on these
images with the same set of parameters as for the original images,
and determined the number of the added stars that are found. We 
estimated the CF as the ratio between the number of artificial stars 
recovered simultaneously in the $V$ and $I$ passbands and the number of 
added stars per one magnitude bin. Table 9 lists the CF values for both 
cluster under study in $V$ band image. For determining the CF 
a number of methods have been described by various authors (cf. Stetson 1987; 
Mateo 1988; Sagar \& Ritchler 1991; Banks et al. 1995) in a CM diagram. 
We adopted the procedure of Sagar \& Ritchler (1991) as this method recovered 
the actual LF better with the mean error of 3 \% up to CF $>$ 0.5 (Mateo 1988).

%
\begin{table}
 \centering
\caption {Variation of completeness factor (CF) in the $V$, $(V-I)$ diagram with 
the MS brightness in both clusters.}
\begin{tabular}{|c|c|c|}
\hline
$V$ mag range&Basel 4&NGC 7067\\
\hline
13 - 14&0.98&0.99\\
14 - 15&0.98&0.99\\
15 - 16&0.97&0.98\\
16 - 17&0.94&0.95\\
17 - 18&0.92&0.93\\
18 - 19&0.90&0.91\\
19 - 20&0.87&0.89\\
\hline
\end{tabular}
\end{table}

\begin{figure}
\psfig{file=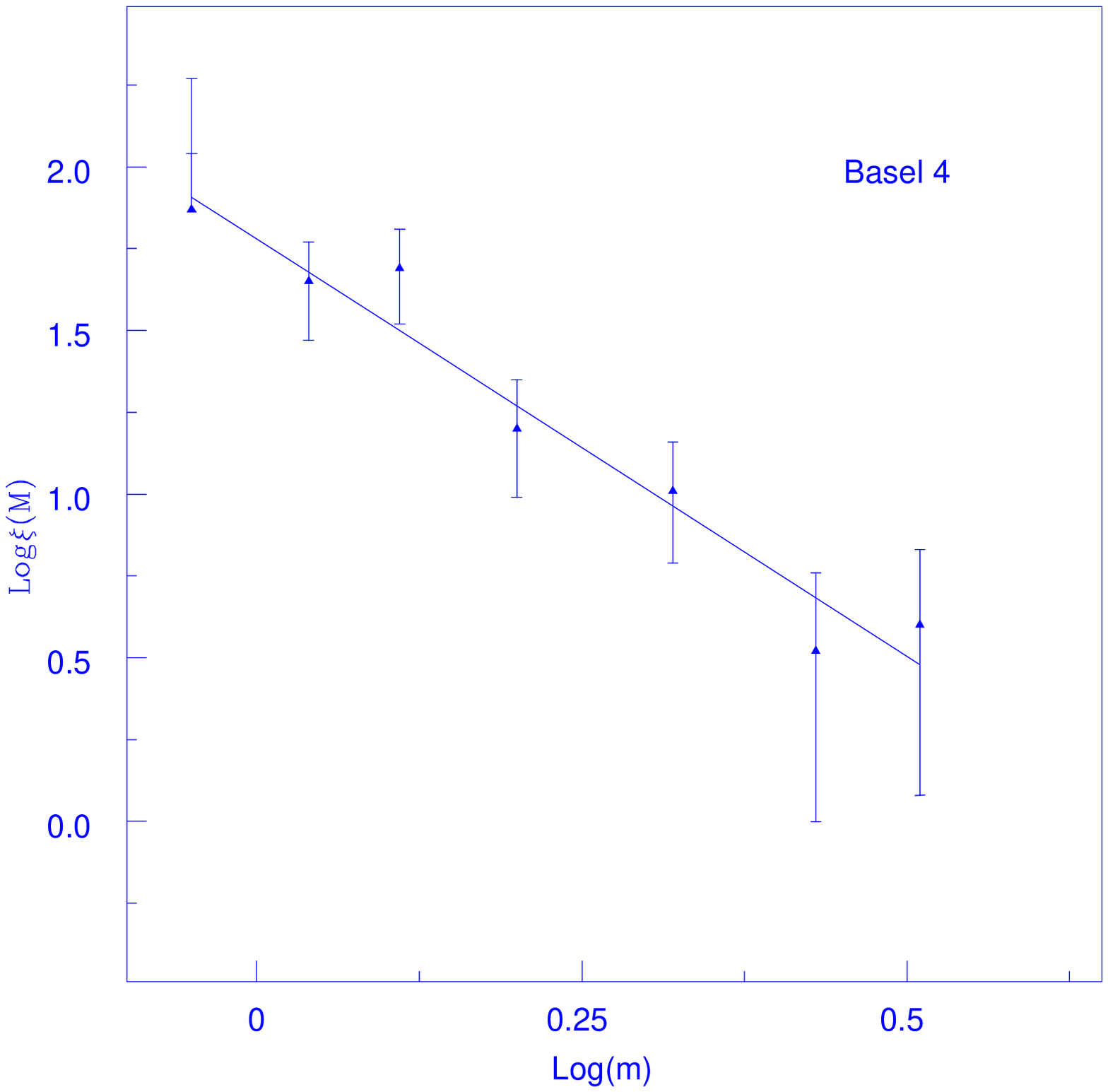 ,width=8.5cm,height=8.8cm}
\vspace{-1.0cm}
\psfig{file=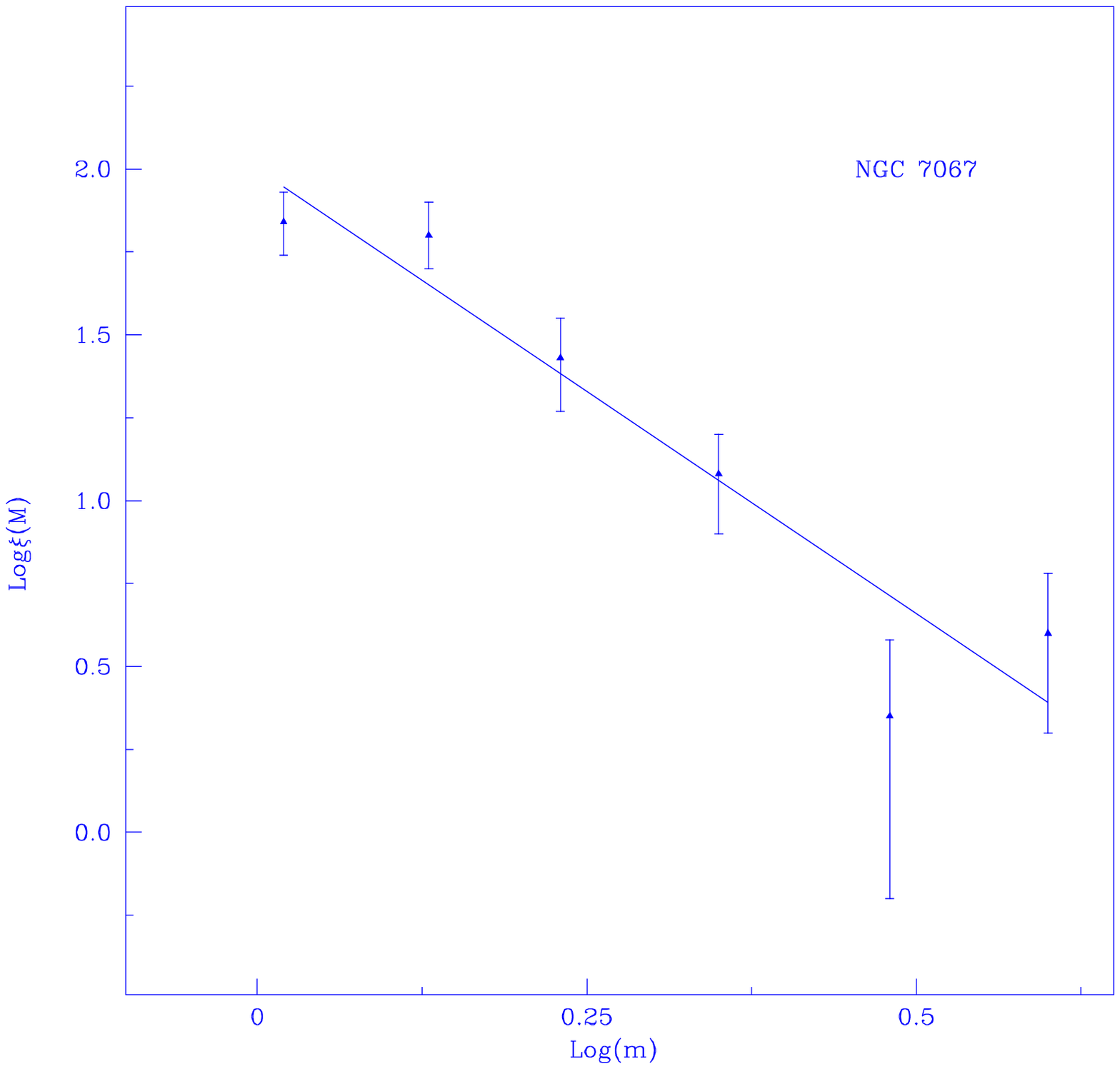 ,width=8.5cm,height=8.8cm}
\caption{The plot shows the mass functions derived using theoretical models of 
Schaerer et al. (1993) for Basel 4 and of Schaller et al. (1992) for NGC 7067.} 
\end{figure}

\subsubsection{Determination of Mass Function}

To construct the luminosity and mass function of the clusters we need to correct 
the luminosity distribution of our selected sample for field star contamination.
To quantify the contamination we adopted the criterion described in Sec. 4.2.  
In $V$, $(V-I)$ CM diagram we defined a strip for the 
MS of cluster region and same strip was also drawn in the CM diagram of 
corresponding field region as shown in Fig 4. Further, we determined the number 
of stars belongs to 
the strip in the cluster region as well as in the field region in each magnitude 
bin. In this way we can estimate the number of field stars
present in various magnitude bins of the cluster region. The observed LFs
 of the cluster and field regions were also corrected for data
incompleteness as well as for differences in area. True LF for the cluster was 
obtained by subtracting the observed LF of field region from the observed LF of 
cluster region.
The MF slope has been derived from the 
mass distribution $\xi$($M$). If $dN$ represents the number of stars 
in a mass bin $dM$ with central mass $M$, then the value of slope $x$ is 
determine from the linear relation 

\vspace{0.2cm}
~~~~~~~~~~~~~~~~log$\frac{dN}{dM}$ = $-$(1+$x$)$\times$log($M$)$+$constant

\vspace{0.2cm}
\noindent using the least-squares solution. The Salpeter (1955) value for the 
slope of MF is $x$ = 1.35.\\

Theoretical models by Schaller et al. (1992) for NGC 7067 and by Schaerer et al. (1993)
for Basel 4 along with the cluster parameters derived by 
us have been used to convert the observed luminosity function to the mass function. 
The plot of MFs of Basel 4 and NGC 7067 is shown in Fig. 11. The value of the MF 
slope along with the mass range and error are given in Table 10, where the quoted 
errors are derived from the linear
least square fit to the data points. Our estimated value of MF slope $x$ is in 
agreement with the Salpeter (1955) value within the error for both the
clusters. However, the error in MF slope is large due to poor statistics of 
cluster members.

\subsection {\bf Mass segregation}
In order to investigate the clusters dynamical evolution and mass segregation 
effect due to energy equipartition, we subdivided the stars into 3 mass range 
3.5$\le$ M$_{\odot}$$<$2.0, 2.0$\le$ M$_{\odot}$$<$1.0 and M$_{\odot}$$<$1.0 
for Basel 4 and 4.5$\le$ M$_{\odot}$$<$2.5, 2.5$\le$ M$_{\odot}$$<$1.0 and 
M$_{\odot}$$<$1.0 for NGC 7067. In Fig. 12 we present cumulative 
radial stellar distribution of stars for different masses. An inspection of 
Fig. 12 shows that both the clusters have mass segregation effect. To check 
whether these mass distribution represent the same kind of distribution or 
not we perform the Kolmogorov-Smirnov (K-S) test. This test shows that mass segregation has taken 
place at confidence level of 65\% for Basel 4 and 70\% for NGC 7067. Further, 
it is important to know whether existing mass segregation is due to dynamical 
evolution or imprint of star formation process.

\begin{figure}
\psfig{file=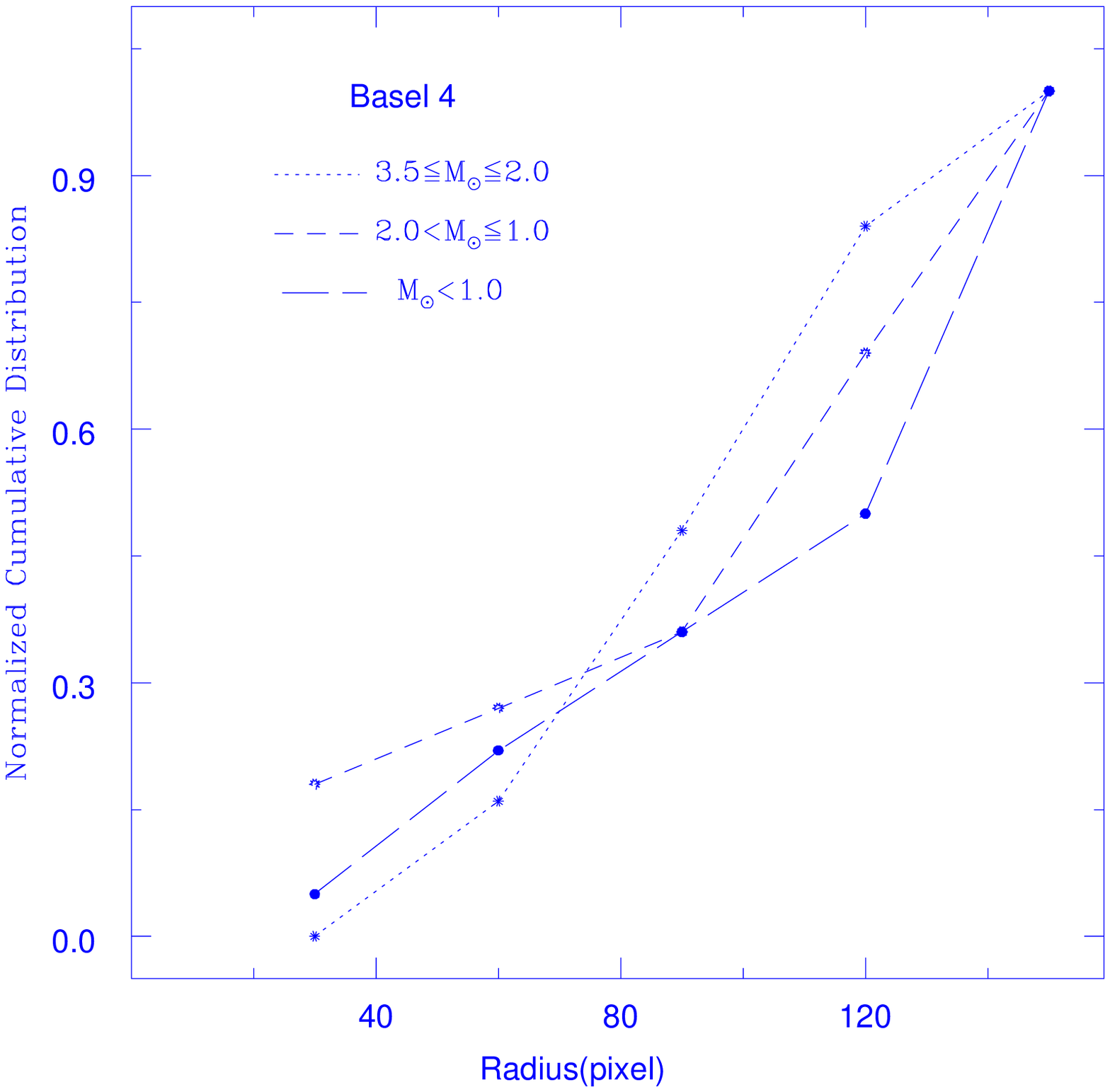 ,width=8.0cm,height=9.8cm}
\vspace{-1.0cm}
\psfig{file=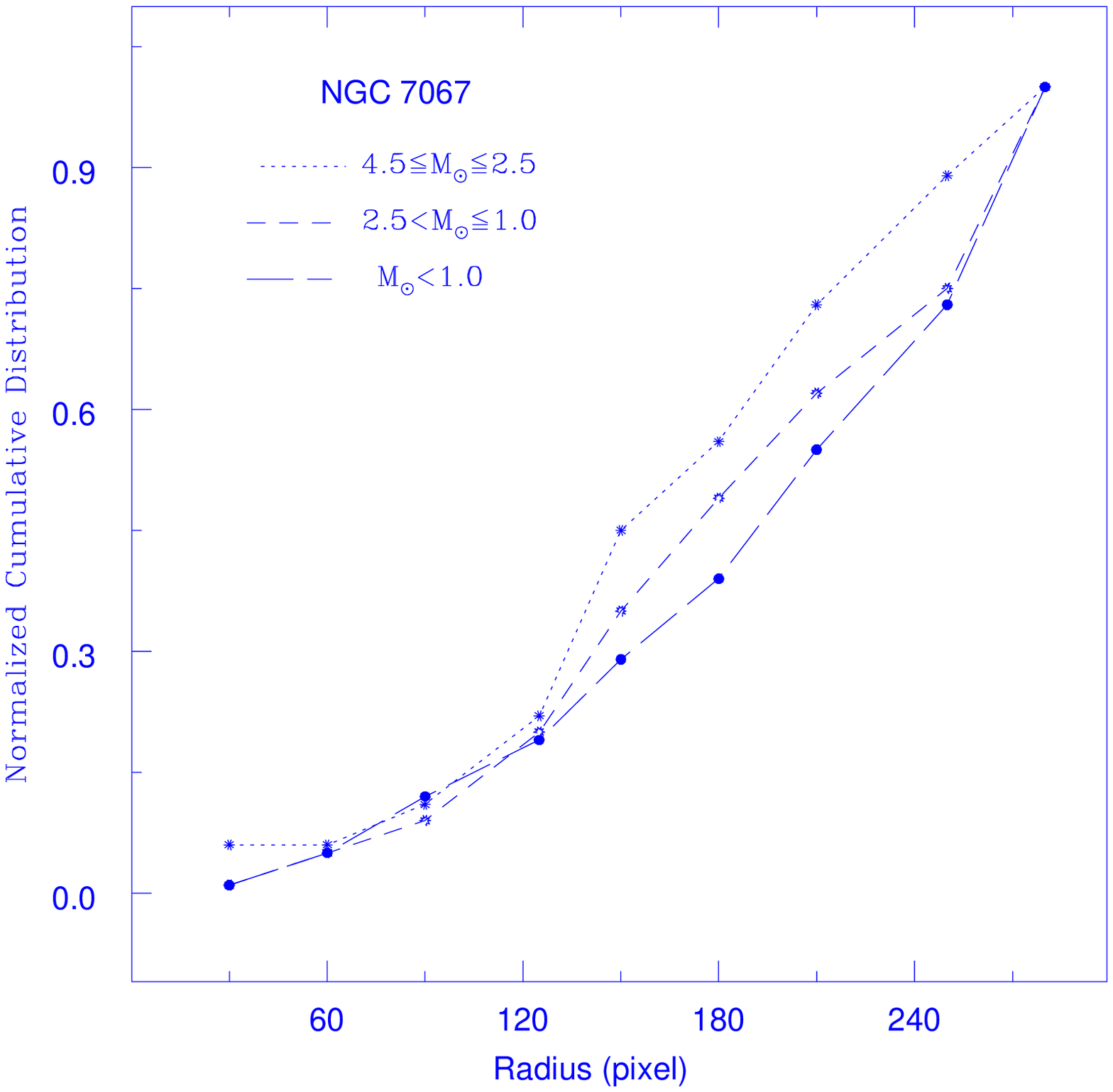 ,width=8.0cm,height=9.8cm}
\caption{Cumulative radial distribution of stars in different mass ranges for 
Basel 4 and NGC 7067.}
\end{figure}

One of the possible cause of mass segregation is the dynamical evolution 
of clusters. Over the lifetime of a star cluster, encounters between its member 
stars gradually lead to an increased degree of energy equipartition throughout 
the cluster. The most important result of this process is that the higher-mass 
cluster members gradually sink towards the cluster center and in the process 
transfer their kinetic energy to the more numerous lower-mass stellar component, 
thus leading to mass segregation. The time scale on which a cluster will have 
lost all traces of its initial conditions is well represented by its relaxation 
time $T_{E}$. It is given by\\
\begin{displaymath}
~~~~~~~~~~~~~~~~~~~~T_{E} = \frac {8.9 \times 10^{5} N^{1/2} R_{h}^{3/2}}{ <m>^{1/2}log(0.4N)}
\end{displaymath}

where $N$ is the number of cluster members, $R$$_{h}$ is the radius containing 
half of the cluster mass and $<m>$ is the mean mass of the cluster stars (cf.
Spitzer \& Hart 1971). The number of probable MS stars is estimated using the 
CM diagrams of the clusters after subtracting the contribution due to field stars
 and applying the necessary corrections for the data incompleteness. For determining 
the R$_{h}$, we assume that the R$_{h}$ is equal to half of the cluster radius 
estimated by us. The angular values are converted to linear values using the 
cluster distances which are derived here. Inclusion of cluster members fainter than the
limiting $V$ magnitude will decrease the value of $<m>$ and increase the value
 of $N$. This will result in higher values of $T_{E}$. Hence the $T_{E}$ values obtained
here may be considered as the lower limit.

\begin{table}
 \centering
\caption{The slope of the mass function derived from LF along with relaxation 
time $T_{E}$.}

\begin{tabular}{cccc}
\hline
cluster&Mass range&Mass Function slope &log T$_{E}$\\
&M$_{\odot}$&($x$)&\\
\hline
Basel 4&0.8 - 3.5&1.55$\pm$0.25&7.0\\
NGC 7067&0.9 - 4.5&1.68$\pm$0.47&7.1\\
\hline
\end{tabular}
\end{table}

A comparison of cluster age with its relaxation time indicates that
the relaxation time is smaller than the age of the clusters. Thus
we can conclude that the clusters under study are dynamically relaxed. It may be 
due to the result of dynamical evolution or imprint of star formation processes
or both.

\section {Conclusions}

In this paper we have presented CCD $UBVRI$ photometry for the stars in the fields of open 
clusters Basel 4 and NGC 7067 for the first time. Using present CCD data in combination with
 2MASS data we derive the following results: \\

(i) Using the $(U-B)$ versus $(B-V)$ colour-colour diagram, we determine the cluster metallicity Z 
$\sim$ 0.008 and 0.02 for Basel 4 and NGC 7067 respectively. The corresponding mean value of 
$E(B-V)$ = 0.45$\pm$0.05 mag and 0.75$\pm$0.05 mag respectively. Interstellar extinction law 
has also been studied using optical as well as near-IR data and found that it is normal in the 
direction of both the clusters. Colour-colour diagram gives the colour excess $E(J-K)$ = 
0.30$\pm$0.20 mag and $E(V-K)$ = 1.60$\pm$0.20 mag for Basel 4 and $E(J-K)$=0.40$\pm$0.20 mag and 
$E(V-K)$ = 2.10$\pm$0.20 mag for NGC 7067.

(ii) Basel 4 and NGC 7067 are located at a distance of 3.0$\pm$0.2 and 3.6$\pm$0.2 Kpc 
respectively. The corresponding ages are 200$\pm$50 and 100$\pm$25 Myrs respectively. They are 
determined by fitting the isochrones of Schaerer et al. (1993) for Z $=$ 0.008 in Basel 4 
and of Schaller et al. (1992) for Z $=$ 0.02 in NGC 7067. Using the 2MASS data we also derived the 
distance and age of both the clusters and they are in agreement with those derived using 
optical data.

(iii) The radial density profiles show that the radius of Basel 4 and NGC 7067 are 1$^{\prime}$.8 
and 3$^{\prime}$.0 respectively which indicate that the clusters under study are compact. At the
cluster distance, they correspond to linear radius of $\sim$ 1.6 and 3.2 pc respectively.

(iv) The values of MF slope are $1.55\pm0.25$ and 1.68$\pm0.47$ for Basel 4 and NGC 7067 respectively. 
They are determined by applying the corrections of data incompleteness and field star contamination 
and are in agreement with the Salpeter (1955) value.

(v) Mass segregation is observed in both Basel 4 and NGC 7067 in the sense that 
massive stars tend to lie near the cluster center. The dynamical relaxation time indicate 
that both the clusters are dynamically relaxed. Thus mass segregation might have occurred 
due to dynamical evolution, or imprint of star formation or both.\\

\section*{Acknowledgments}

We thank the referee for valuable comments which have improved the quality of this paper. 
We are grateful to Dr. Vijay Mohan for helping in data reduction. This study made use of 
2MASS and WEBDA.

\bsp
\label{lastpage}

\end{document}